# Prior-Adapted Progressive Time-Resolved CBCT Reconstruction Using a Dynamic Reconstruction and Motion Estimation Method

Ruizhi Zuo[1], Hua-Chieh Shao[1], You Zhang[1]

[1]*The Medical Artificial Intelligence and Automation (MAIA) Laboratory*
*Department of Radiation Oncology, University of Texas Southwestern Medical Center, Dallas, TX 75390, USA*

Corresponding address:

You Zhang
Department of Radiation Oncology
University of Texas Southwestern Medical Center
2280 Inwood Road
Dallas, TX 75390
Email: You.Zhang@UTSouthwestern.edu
Tel: (214) 645-2699

## Abstract

**Background:**

Cone-beam CT (CBCT) captures on-board volumetric anatomy for image guidance and treatment adaptation in radiotherapy. To compensate for respiration-induced anatomical motion, time-resolved CBCT is highly desired to capture the spatiotemporal anatomical variations but faces challenges in accuracy and efficiency due to substantial optimization needed in image reconstruction and motion modeling.

**Purpose:**

We proposed a fast time-resolved CBCT reconstruction framework, based on a dynamic reconstruction and motion estimation method with new reconstructions initialized and conditioned on prior reconstructions in an adaptive fashion (DREME-adapt).

**Materials and Methods:**

DREME-adapt reconstructs a time-resolved CBCT sequence from a fractional standard CBCT scan while simultaneously generating a machine learning-based motion model that allows single-projection-driven intra-treatment CBCT estimation and motion tracking. Via DREME-adapt, a virtual fraction is generated from a pre-treatment 4D-CT set of each patient for a clean, 'cold-start' reconstruction. For subsequent fractions of the same patient, DREME-adapt uses pre-derived motion models and reference CBCTs as initializations to drive a 'warm-start' reconstruction, based on a lower-cost refining strategy. Three strategies: DREME-cs which drops the 'warm-start' component, DREME-adapt-vfx which uses a fixed initialization (virtual fraction's reconstruction results), and DREME-adapt-pro which initialize reconstructions through a progressive daisy chain scheme (virtual fraction for fraction 1, fraction 1 for fraction 2, and so on), were evaluated on a digital phantom study (7 motion/anatomical scenarios) and a patient study (7 patients).

**Results:**

DREME-adapt allows fast and accurate time-resolved CBCT reconstruction. For the XCAT simulation study, DREME-adapt-pro achieves image reconstruction relative error of 0.14±0.01 and tumor center-of-mass tracking error of 0.92±0.62 mm (mean±s.d.), compared to 0.15±0.01 and 1.06±0.73 mm for DREME-adapt-vfx, and 0.18±0.01 and 1.96±1.35 mm for DREME-cs. For the real-time motion inference test dataset of the patient study, DREME-adapt-pro localizes moving lung landmarks to a mean±s.d. error of 2.21±1.79 mm. In comparison, the corresponding values for DREME-adapt-vfx and DREME-cs were 2.53±1.93 mm and 3.22±2.88 mm, respectively. The DREME-adapt-pro training takes 11 minutes, only 15% of the original DREME algorithm.

**Conclusions:**

With high efficiency and accuracy, DREME-adapt-pro allows on-board time-resolved CBCT reconstruction and enhances the clinical adoption potential of the DREME framework.

## 1. Introduction

Cone-beam computed tomography (CBCT) is widely used in clinical practice, offering on-board high spatial resolution volumetric imaging guidance for treatment setup, dose verification, and adaptive radiotherapy[1,2]. With the development of CBCT systems, a typical scan takes about 6-60 seconds for a full 360° acquisition[3,4]. However, patient anatomical motion, mainly due to respiration with a cycle of 3–5 seconds, can introduce motion artifacts and blurring in the reconstructed images[5,6]. Respiratory motion in the thoracic and abdominal regions leads to uncertainties in target shape and position, compromising the precision of radiotherapy. To address this concern and resolve the underlying motion, four-dimensional (4D) CBCT was developed as the current clinical solution. 4D-CBCT organizes the projections into predefined respiratory phases and assembles semi-static CBCTs reconstructed from each bin into a sequence of respiration-resolved 3D CBCT images. This enables visualization of tumor motion throughout the breathing cycle, supports accurate internal target volume (ITV) definition, and promotes safe and effective radiation delivery. However, achieving high 4D-CBCT image quality is challenging, as the phase-sorted data is inherently sparse and often contains large angular gaps, leading to undersampling artifacts such as streaks[7]. Iterative methods, especially those combined with compressed sensing[8], were proposed to suppress the undersampling artifacts by enforcing consistency with the measured projections and sparsity in a feature domain(gradient of the image, for instance), typically through regularizations like the total variation. However, their efficacy is limited, especially when sorted projections are highly limited. Another approach is to build a motion model from prior images, such as a patient-specific 4DCT[9,10], to achieve dimension reduction and enable few projection-based, deformation-driven image reconstruction. However, the deformation-based approach faces the challenge of the intrinsic differences in motion patterns, anatomy, and intensities between the prior and target images[11]. More importantly, the motion sorting/binning of conventional 4D-CBCT assumes that the patient motion is regular and periodic, which is generally false and limits its ability to capture time-resolved irregular motion[12]. To resolve both regular and irregular respiratory-dominated motion to guide radiotherapy delivery, time-resolved CBCT imaging, by which each X-ray projection is reconstructed into its own CBCT image, is highly desired. Time-resolved dynamic CBCT imaging provides the highest spatiotemporal resolution that can be achieved with an X-ray scanner but is difficult for conventional reconstruction methods due to extreme undersampling. Without using the prior knowledge, image-based motion estimation can be achieved by enforcing of the consistency of the CBCT sinogram's Fourier properties[13] or epipolar symmetry[14]. However, the consistency-based methods present limited sensitivity to some motion components, and are limited to a field-of-view constrained to the vicinity of the central slice, beyond which most consistency conditions are not valid. Autofocus-based methods have been demonstrated for rigid CBCT motion estimation[15] and soft-tissue abdominal deformable motion compensation[16]. However, the requirement of motion-free CBCT volumes for training is difficult to meet for clinical CBCT datasets.

In the past few years, deep learning (DL)-based approaches have become popular in CBCT reconstruction because of the low inference latency and improved accuracy[17-21]. DL solutions have demonstrated the feasibility of 3D image reconstruction from highly sparse 2D X-ray projections[22-24]. Ying et al.[18] proposed reconstructing a CBCT from two orthogonal X-rays using a generative adversarial network (GAN). The generator consists of 2D encoders, a fusion module, and a 3D decoder to generate the 3D volume. Although the GAN-based model can generate high-quality images, the fine details and structure may not be reliable due to the synthetic nature of the GAN models. Shen et al.[25] introduced a network integrating geometric priors of the imaging system to bridge 2D and 3D image space to assist CBCT image reconstruction from ultra-sparse projections. However, the sparse measurements did not provide sufficient information and the DL methods have to rely on strong priors for precise reconstruction, resulting in poor generalization across diverse datasets. Additionally, the lack of 'ground truth' clinical CBCT datasets may further limit the potential of such supervised learning, which usually requires a large training dataset.

Implicit neural representation (INR)[26] learning has emerged as a novel machine learning technique to map spatial coordinates to corresponding physical quantities (for instance, CBCT image intensities), serving a universal function approximator that enables high-resolution and memory-efficient representations of volumetric images. Directly representing image volumes as neural networks (rather than gridded voxels), INR-based strategy allows the use of deep learning frameworks to efficiently and effectively optimize the image-representing networks for iterative, limited-sample-based image reconstruction. Compared with direct voxel reconstruction, INR models the CBCT volume as a continuous function that can be sampled at arbitrary resolutions with lower memory cost than a full voxel array[26]. The inductive bias of INR[27] helps to suppress high-frequency artifacts and improve stability when reconstructing CBCTs from limited-view projections, resulting in reconstructions that are anatomically consistent and less prone to artifacts. Based on INR, the simultaneous Spatial and Temporal Implicit Neural Representation (STINR) framework[28] is developed to address the time-resolved dynamic CBCT reconstruction challenge. It decouples the time-resolved dynamic CBCT reconstruction into solving a spatial INR to represent a reference static CBCT image, and temporal INRs to learn intra-scan, time-resolved dynamic motion from each X-ray projection. The solved dynamic motion, after being applied to the reference CBCT, yields projection-specific dynamic CBCT images. Such a motion-compensated reconstruction strategy uses all available projections via a 'one-shot' learning and self-supervised framework to simultaneously optimize the spatial and temporal INRs. It effectively addresses the undersampling challenge to reconstruct a high-quality dynamic CBCT set without the need of population-based large-scale training. Based on STINR, Shao et al. further improves the reconstruction strategy by removing its use of a fixed prior motion model, yielding a prior-model-free STINR (PMF-STINR) technique. PMF-STINR learns an on-the-fly motion model during the reconstruction, the efficacy of which has been validated on multi-institutional real patient scans[29]. Recently, on the basis of PMF-STINR, Shao et al. further proposed a joint dynamic reconstruction and motion estimation (DREME) framework for real-time CBCT imaging and motion tracking[30]. In addition to reconstructing a dynamic CBCT set and deriving a motion model from a pre-treatment 3D CBCT scan, DREME simultaneously learns a convolutional neural network (CNN)-based motion encoder, which can infer motion model coefficients from arbitrarily angled, singular X-ray projections to derive real-time deformable motion. The dual-task learning system of DREME enables real-time 3D deformable motion and CBCT inference from later-acquired intra-treatment X-ray projections for continuous motion monitoring and tracking. While achieving high accuracy in dynamic reconstruction and real-time motion tracking, the training time for DREME is approximately 3-4 hours (based on CBCTs of 200×200×100 voxels) on an Nvidia V100 GPU card (~73 mins on Nvidia RTX 4090). The time-consuming nature of DREME, driven by prolonged iterations for from-scratch 'one-shot' motion modeling and image reconstruction, hinders its clinical practicality when on-the-fly reconstruction is needed at each treatment fraction.

To overcome the efficiency limitations of DREME reconstruction, in this study we developed a prior-adapted reconstruction strategy based on the DREME framework (DREME-adapt). By DREME-adapt, patient-specific pre-treatment 4D-CT images are used to simulate motion-involved cone-beam projections to mimic a 'virtual' treatment fraction, and the projections are trained by the original DREME framework to yield a virtual-fraction DREME model. Because patients usually exhibit similar anatomy and motion patterns throughout the treatment course. DREME-adapt then uses the reference CBCT and the motion model from the virtual-fraction DREME reconstruction as 'warm-start' initializations, to jump-start and speed-up the dynamic CBCT reconstructions at real treatment fractions. Under the framework of DREME-adapt, we investigated two fine-tuning strategies: 1. DREME-adapt-vfx, which always uses the virtual-fraction DREME model to warm-start the CBCT reconstruction at each treatment fraction; 2. DREME-adapt-pro, which initializes the 1st treatment fraction with the virtual-fraction DREME model. The DREME model derived from the 1st treatment fraction is then applied to initialize the reconstruction of the 2nd treatment fraction, forming a progressive daisy chain. In addition, we also evaluated the original DREME framework without the 'warm-start' component for comparison (DREME-cs, 'cold-start'). All strategies were evaluated with both a digital phantom simulation study and a lung patient dataset study. Compared with the original DREME reconstruction, DREME-adapt-pro reduces the reconstruction time by 85% (~11 mins on RTX 4090) while preserving high reconstruction accuracy. During the adaptive radiotherapy, DREME-adapt-pro can rapidly train a patient- and fraction-specific model to enable dynamic and real-time CBCT imaging, to track tumor motion during treatment with high accuracy and image quality, thereby potentially allowing a safe reduction of the planning target volume margin[31] and reducing the treatment toxicity[32]. By monitoring the intra-fraction anatomical variation, DREME-adapt-pro could help dynamically reoptimize the treatment plan to accommodate the change of patient motion/anatomy, which provides an opportunity to safe dose escalation[33], as well as precise dose accumulation[34] and treatment analysis[35].

## 2. Materials and Methods

DREME-adapt reconstructs a time-resolved CBCT sequence from a fractional standard CBCT scan while simultaneously generating a machine-learning-based motion model that allows single-projection-driven intra-treatment CBCT estimation and motion tracking. Via DREME-adapt, we use the pre-treatment 4D-CT scan of a patient to simulate motion-involved cone-beam projections for a clean 'cold-start' reconstruction. For subsequent real treatment fractions of the same patient, DREME-adapt uses pre-derived motion models and reference CBCTs as initializations to drive 'warm-start' reconstructions, based on a lower-cost refining strategy.

2.1 Dynamic and real-time CBCT reconstruction by the original DREME framework

In general, the original DREME framework reconstructs a dynamic sequence of 3D CBCT volumes from a pre-treatment projection set $\boldsymbol{p}$. Due to the high spatiotemporal correlations within a dynamic CBCT set, the dynamic anatomy could be decomposed as a static reference anatomy $\boldsymbol{I}_{ref}(\boldsymbol{x})$ and a time-varying, projection-dependent deformable vector field (DVF) set $\boldsymbol{d}(\boldsymbol{x},p)$ on the reference anatomy, i.e.,

$$\boldsymbol{I}(\boldsymbol{x},p) = \boldsymbol{I}_{ref}\big(\boldsymbol{x} + \boldsymbol{d}(\boldsymbol{x},p)\big), \tag{1}$$

where $\boldsymbol{x}$ denotes the voxel coordinates and $p \in \boldsymbol{p}$. Since solving time-dependent DVFs remains an ill-posed spatiotemporal inverse problem, a principal component analysis (PCA)-based spatiotemporal decoupling has demonstrated an effective representation of the dynamic lung motion[36]. Under a similar philosophy, we decomposed DVFs as a summation of products between projection-specific temporal components ($\boldsymbol{w}_i(p)$) and projection-agnostic spatial components ($\boldsymbol{e}_i(\boldsymbol{x})$), rendering a motion model:

$$\boldsymbol{d}(\boldsymbol{x},p) = \sum_{i=1}^{3} \boldsymbol{w}_i(p) \times \boldsymbol{e}_i(\boldsymbol{x})\,. \tag{2}$$

By DREME, the reference anatomy $\boldsymbol{I}_{ref}(\boldsymbol{x})$ is represented by a spatial INR that applies mapping $\boldsymbol{x} \mapsto \boldsymbol{I}_{ref}(\boldsymbol{x})$ from a voxel coordinate $\boldsymbol{x}$ to the corresponding attenuation coefficient $\boldsymbol{I}_{ref}(\boldsymbol{x})$. $\boldsymbol{w}_i(p)$ is represented by temporal motion coefficients derived by a CNN-based motion encoder, and $\boldsymbol{e}_i(\boldsymbol{x})$, represented by a B-spline interpolant, captures the spatial motion variations that compose a basis set (motion basis components, MBCs). Due to the redundancy and correlation of time-dependent DVFs, three levels ($i = 1,2,3$) of the spatial decomposition are sufficient to represent the dynamic lung motion[29,30]. Although using more levels could provide additional flexibility, it also increases the risk of overfitting. The spatial INR, the CNN-based motion encoder, and the MBCs are all learned on-the-fly directly from the raw cone-beam projections of a pre-treatment CBCT scan, via a 'one-shot' and self-supervised learning scheme. The workflow and the network architecture of the DREME framework is summarized in Fig. 1. DREME has two learning tasks. The first learning task (Fig. 1, upper panel) is dynamic CBCT reconstruction, where acquired cone-beam projections serve as input to generate a dynamic CBCT sequence. All of $\boldsymbol{I}_{ref}(\boldsymbol{x})$, MBC components $\boldsymbol{e}_i(\boldsymbol{x})$, and their scores $\boldsymbol{w}_i(p)$ are updated through a multi-stage coarse-to-fine training strategy with projection-based similarity loss and additional regularization losses. The second learning task (Fig. 1, middle panel) is to further train the CNN motion encoder to be robust to different motion scenarios (especially those not contained in the acquired projections) for accurate real-time motion estimation. The CNN encoder is trained to extract the temporal coefficients of the motion basis components from the pre-treatment CBCT projections. Additionally, it also serves to map future, intra-treatment projections into real-time temporal coefficient to enable real-time on-board CBCT inference. Since the $I_{ref}(x)$ and MBC components $\boldsymbol{e}_i(\boldsymbol{x})$ become stable near the end of training, they are frozen then to focus on the training of the CNN encoder to improve its generalizability. The on-board real-time CBCT inference stage (Fig. 1, lower panel), the intra-treatment cone-beam projection $p$ is fed into the network to infer the MBC scores and the corresponding real-time CBCT and DVFs. If $\boldsymbol{I}_{ref}$ here is replaced by the contour of a tracking target, DREME can be used for real-time target localization.

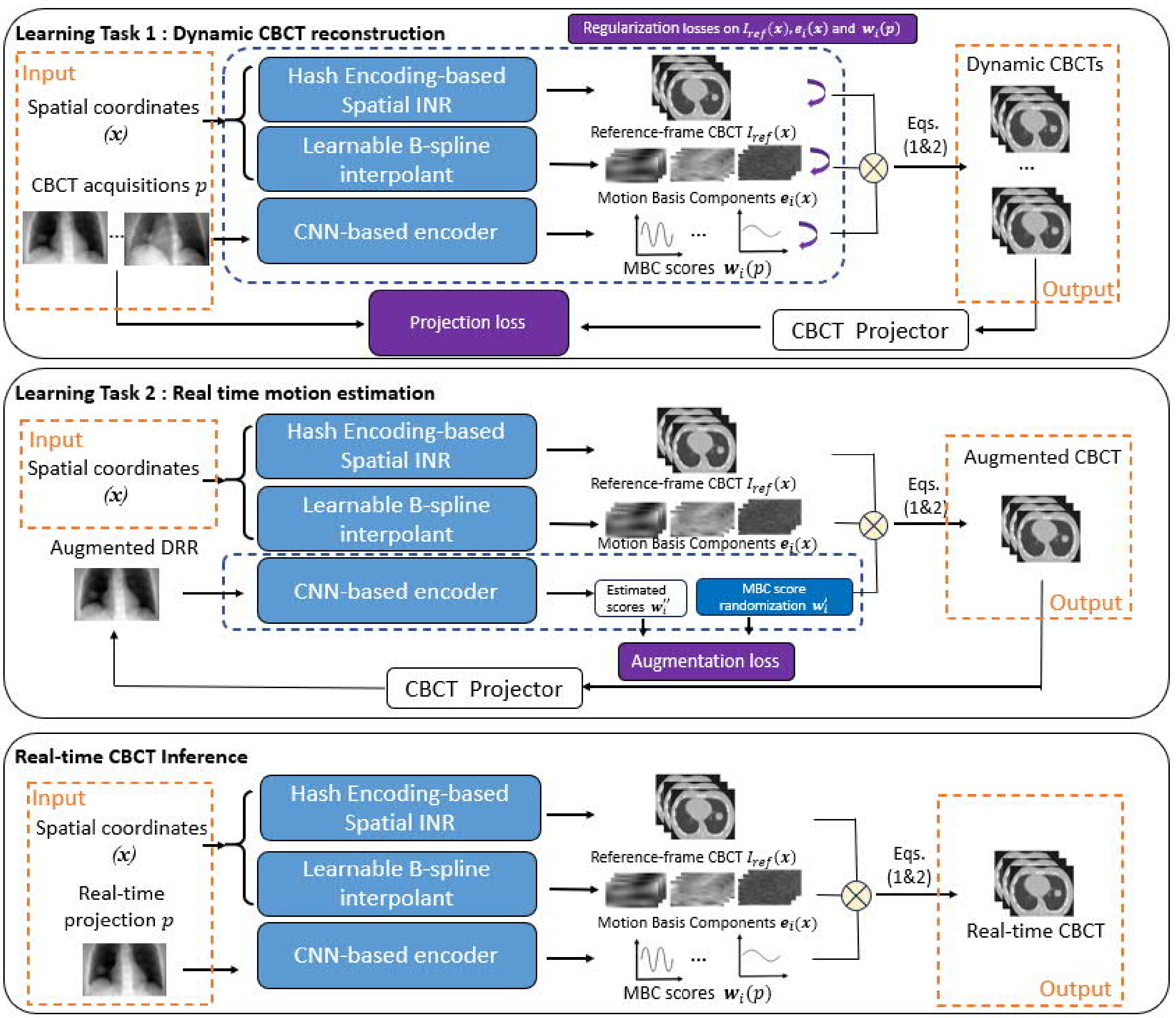

**Figure 1**. The DREME framework for dynamic CBCT reconstruction and real-time motion estimation. In learning task 1 (*upper panel*), a spatial implicit neural representation (INR) network represents the reference CBCT $\boldsymbol{I}_{ref}(\boldsymbol{x})$. The learnable B-spline interpolant and the CNN motion encoder represent the spatial $\boldsymbol{e}_i(\boldsymbol{x})$ and temporal components $\boldsymbol{w}_i(p)$ of the deformation vector fields $\boldsymbol{d}(\boldsymbol{x}, p)$, respectively. From the three components, time-resolved dynamic CBCTs are derived using Eqs. 1 and 2. The reconstruction is driven by the projection loss between the CBCT acquisition $\boldsymbol{p}$ and digitally reconstructed radiographs (DRRs) projected from the derived dynamic CBCT sequence. Different regularization losses are incorporated into the training to address the ill-posed nature of the spatiotemporal inverse problem. The learnable parameters enclosed within the blue dashed box are subject to update during the training. In learning task 2 (*middle panel*), deformable motion augmentation is introduced by randomizing the temporal coefficients of the motion model ($\boldsymbol{w}_i'$) to sample random motion. The projection angle augmentation is also implemented to simulate DRRs at random projection angles, to make the CNN motion encoder robust to arbitrary projection directions. The real-time CBCT inference stage (*lower panel*) takes an on-board cone-beam projection $p$ from an arbitrary angle as input, to estimate the MBC scores, the corresponding motion field, and the real-time CBCT/contour.

2.2 Loss function design and training strategy of DREME

DREME uses a progressive training strategy with increasing training complexity to improve the learning efficiency and avoid overfitting. It consists of a low-resolution stage (stage I) with an isotropic 4×4×4 $mm^3$ reconstruction resolution and a high-resolution stage (stage II) with a 2×2×2 $mm^3$ resolution, each containing multiple substages, with distinct total loss functions combining differently weighted loss components.

During the first training substage (I-a), the spatial INR model is initialized using a motion-blurred, approximate CBCT $\boldsymbol{I}_{app}(\boldsymbol{x})$, reconstructed using all available projections by the Feldkamp-Davis-Kress (FDK) method[37]. The image-domain similarity loss $L_{sim}^{im}$ between $\boldsymbol{I}_{ref}(\boldsymbol{x})$, which is voxelized from the spatial INR, and $\boldsymbol{I}_{app}(\boldsymbol{x})$ is calculated as:

$$L_{sim}^{im} = \frac{1}{N_{voxel}} \sum_{l=1}^{N_{voxel}} \left| \boldsymbol{I}_{ref}(\boldsymbol{x}) - \boldsymbol{I}_{app}(\boldsymbol{x}) \right|^2, \tag{3}$$

where $N_{voxel}$ denotes the number of voxels in the CBCT. This substage is mainly to provide the INR with a warm start, allowing the following optimizations to stabilize and converge faster. In the next substage I-b, the spatial INR is then optimized with a projection domain similarity loss $L_{sim}^{prj}$ between the digitally reconstructed radiographs (DRRs) generated from the reference CBCT $\boldsymbol{I}_{ref}(\boldsymbol{x})$ by the projector $\mathcal{P}$ and the corresponding cone-beam projections $\boldsymbol{p}$:

$$L_{sim}^{prj}(\boldsymbol{I}_{ref}(\boldsymbol{x})) = \frac{1}{N_{batch}\, N_{pixel}} \sum_{N_{batch}} \sum_{N_{pixel}} \left| \mathcal{P}[\boldsymbol{I}_{ref}(\boldsymbol{x})] - \boldsymbol{p} \right|^2, \tag{4}$$

where $N_{batch}$ and $N_{pixel}$ are the number of projection samples in a training batch and the number of pixels in a projection, respectively. In addition to the similarity loss, DREME incorporates the total variation (TV) [38] regularization into the CBCT reconstruction to suppress high-frequency image noise while preserving anatomical edges:

$$L_{TV} = \frac{1}{N_{voxel}} \sum_{N_{voxel}} \left| \nabla \boldsymbol{I}_{ref}(\boldsymbol{x}) \right|, \tag{5}$$

where $\nabla$ denotes the gradient operator. In the first two substages, a quick FDK reconstruction with additional projection-based iterative reconstruction is an economical solution without incurring much additional memory/time cost. Some more advanced algorithms, such as model-based iterative reconstruction (MBIR)[39] can generate higher-quality images than FDK. However, it is computationally intensive due to iterative updates over all voxels. Replacing the initialization with MBIR may incur additional memory demand and significantly increase the computation time. Since motion modeling is not performed in the initialization stages, the image quality gain from the MBIR reconstruction may also be diminished due to the motion blurriness and artifacts intrinsic to the reconstructed volume, thus a simpler initialization strategy was selected for DREME.

In the substage I-c, DREME added the motion model into training, using different regularization losses on the MBCs $\boldsymbol{e}_i(x)$ and MBC scores $\boldsymbol{w}_i(p)$. For the MBCs ($\boldsymbol{e}_i(x)$), it enforced their ortho-normality as:

$$L_{MBC} = \frac{1}{9} \sum_{k=x,y,z} \sum_{i=1}^{3} \left( \left| \left\| e_{i,k} \right\|^2 - 1 \right|^2 + \sum_{j=i+1}^{3} \left| e_{i,k} \cdot e_{j,k} \right|^2 \right), \tag{6}$$

where, the subscript $i = 1,2,3$ and $k = x, y, z$ represent the MBC level index, and Cartesian components index, respectively. The normalization removes the ambiguity in the spatiotemporal decomposition of the low-rank motion model, and the orthogonality constraint eliminates redundancy among the MBCs across different levels of the B-spline interpolant. The MBC scores $\boldsymbol{w}_i(p)$ are regularized with a zero-mean score loss to remove the constant non-zero baseline:

$$L_{zms} = \frac{1}{9} \sum_{k=x,y,z} \sum_{i=1}^{3} \left| \frac{1}{N_p} \sum_{p} w_{i,k}(p) \right|, \quad (7)$$

where $N_p$ denotes the total number of projections.

In addition, A DVF self-consistency regularization loss is included. This regularization ensures that applying the inverse DVFs $\boldsymbol{d}^{-1}(\boldsymbol{x}, p)$ to the deformed (dynamic) CBCTs $\boldsymbol{I}(\boldsymbol{x}, p)$ brings them back to the reference CBCT $\boldsymbol{I}_{ref}(\boldsymbol{x})$, thus helping address the ill-posed nature of the reconstruction problem:

$$L_{SC} = \frac{1}{N_{batch}} \sum_{N_{batch}} \frac{1}{N_{voxel}} \sum_{N_{voxel}} \left| \boldsymbol{I}'(\boldsymbol{x}, p) - \boldsymbol{I}_{ref}(\boldsymbol{x}) \right|^2, \quad (8)$$

where $\boldsymbol{I}'(\boldsymbol{x}, p) = \boldsymbol{I}(\boldsymbol{x} + \boldsymbol{d}^{-1}(\boldsymbol{x}, p), p)$. The inverse DVFs were calculated using a fixed-point iterative algorithm with three iterations [40]. In this stage, the projection similarity loss $L_{sim}^{prj}(\boldsymbol{I}(x, p))$ is defined on the dynamic CBCTs $\boldsymbol{I}(x, p)$:

$$L_{sim}^{prj}(\boldsymbol{I}(\boldsymbol{x}, p)) = \frac{1}{N_{batch}\, N_{pixel}} \sum_{N_{batch}} \sum_{N_{pixel}} |\mathcal{P}[\, \boldsymbol{I}(\boldsymbol{x}, p)] - \boldsymbol{p}|^2, \quad (9)$$

In training stage II, the reference CBCT $\boldsymbol{I}_{ref}$ is first 2x upsampled to accommodate the high-resolution training. A 2x denser sampling of spatial coordinate was performed for input to the INR for high-resolution training. DREME follows the strategies and loss functions used in stage I to form substages II-a, II-b, and II-c. Substage II-a and II-b are designed to progressively enhance the spatial INR from low to high spatial resolution, whereas substages II-c jointly optimize the CNN motion encode, B-spline interpolator, and the spatial INR for more accurate and coherent spatiotemporal reconstruction. Additionally, to improve the robustness of the CNN motion encoder to varying projection angles, angular augmentation is added to substages II-b and II-c, by synthesizing DRRs at random angles from the solved dynamics CBCTs. The MBC scores inferred by the CNN motion encoder from these DRRs were compared with the known scores to calculate the angular augmentation loss. Last, a substage II-d is added to enable simultaneous angular and motion augmentation training. During this substage, the spatial INR and motion basis components (B-spline interpolant) are frozen to focus on improving the generalizability of the CNN motion encoder. Based on augmented MBC scores $w'_{i,k}(p)$corresponding DVFs (Eq. 2) and deformation-augmented CBCTs (Eq. 1) are obtained. Random-angled DRRs are generated from these deformation-augmented CBCTs and fed into the motion model to infer the MBC scores $w''_{i,k}(p)$. The augmentation loss is defined as the sum of squared errors between the known scores $w'_{i,k}(p)$ and CNN-predicted scores $w''_{i,k}(p)$ to train a motion- and projection angle-robust CNN for real-time motion prediction:

$$L_{AUG} = \frac{1}{9N_{batch}} \sum_{N_{batch}} \sum_{k=x,y,z} \sum_{i=1}^{3} \left| w''_{i,k}(p) - w'_{i,k}(p) \right|^2. \quad (10)$$

For the original DREME framework, the training epochs for each training stage (from I-a to II-d) are 400, 700, 1700, 1000, 1000, 1000, and 1000. Detailed learning rates and loss function weightings can be found in Table 1 of the corresponding publication[30]. The original DREME framework learns the reference CBCT and the motion model (including both the CNN motion encoder and B-spline-based MBCs) from scratch each time with a new set of CBCT projections, which is time consuming. However, different treatment fractions of the same patient usually share very similar anatomy and motion models. Thus, faster reconstructions of new treatment fractions could be achieved by initializing and warming up with prior reconstruction results in an adaptive manner. Based on this, we developed DREME-adapt (with two variants) and compared them with a 'cold-start' version of DREME reconstruction (DREME-cs), with details laid out in the following.

2.3 DREME-adapt workflow

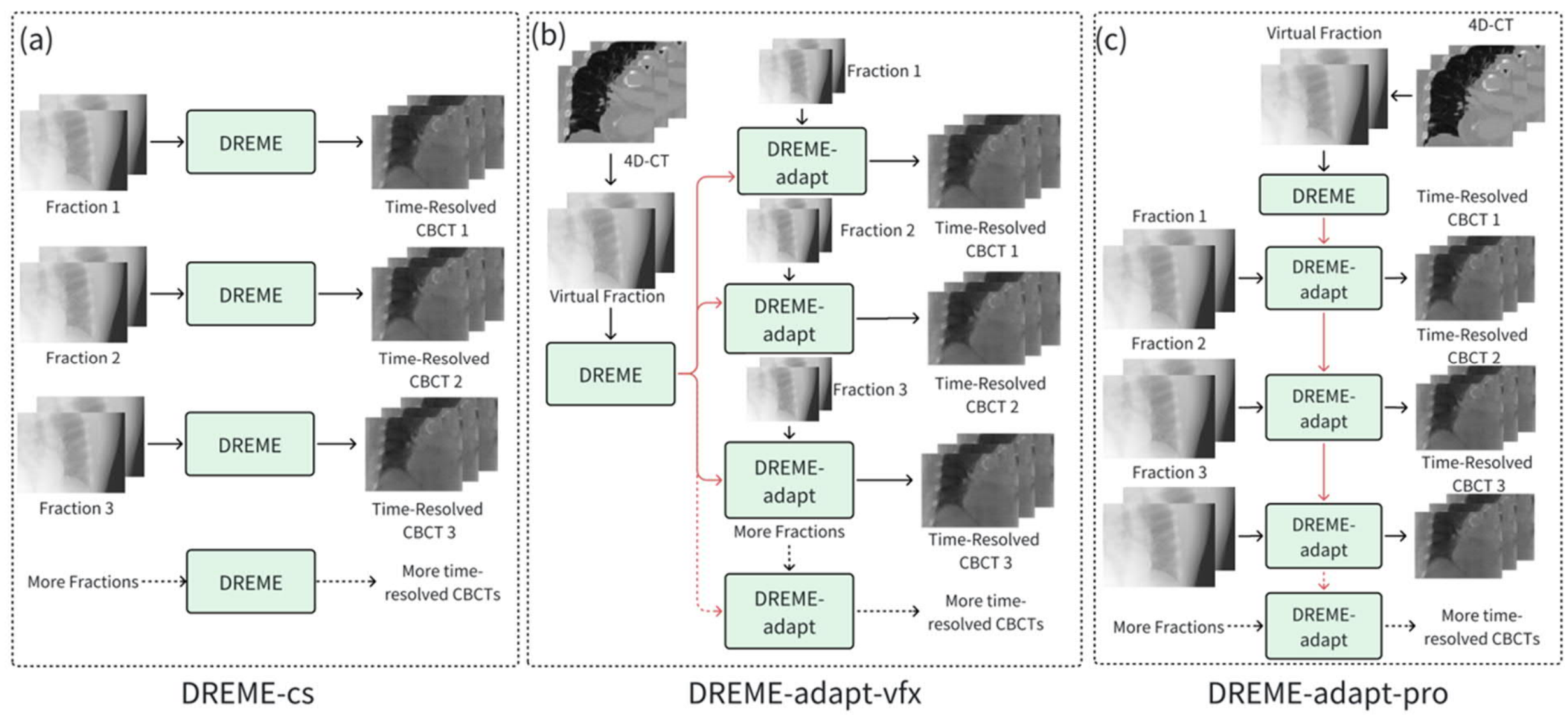

**Figure 2**. The workflow of three different training strategies. The virtual fraction is generated from pre-treatment 4D-CT scans. The red arrows indicate DREME-adapt reconstructions initialized with the spatial INR (reference CBCT $\boldsymbol{I}_{ref}$) and the motion model (B-spline interpolant, $\boldsymbol{e}_i(\boldsymbol{x})$; and CNN-based motion encoder, $\boldsymbol{w}_i(p)$) derived from previous fractions. In (a) DREME-cs, the DREME model of each fraction is trained separately from scratch. (b) DREME-adapt-vfx uses the reconstruction outputs of the 4D-CT-simulated virtual fraction as a fixed set to initialize all the subsequent reconstructions. (c) DREME-adapt-pro is a progressive, inter-fractional strategy. For each fraction, the DREME model was progressively fine-tuned from the DREME model learned from the preceding fraction (for fraction 1, it uses the DREME model derived from the 4D-CT-simulated virtual fraction) with new scan projections, forming a daisy chain.

Figure 2 illustrates the workflow of three different fractional training strategies. DREME-cs follows the original DREME framework in a 'cold-start' manner, with each fractional model trained individually and from scratch using the daily CBCT scan data. The two DREME-adapt variants, on the other hand, try to use the patient anatomy and motion correlations between the simulation and the treatment, or between different treatment fractions to greatly accelerate the training speed. For DREME-adapt-vfx, a virtual fraction of CBCT imaging is introduced, which can be simulated from a pre-treatment 4D-CT scan. The virtual fraction is trained using the original DREME model and training setup. The resultant DREME model of the virtual fraction is then used to initialize the dynamic CBCT reconstructions of all sequential treatment fractions. Considering the potential gradual change of patient anatomy and motion patterns during the treatment, the DREME-adapt model derived from the virtual fraction may become less and less relevant in later treatment fractions. To maximize the relevance and utility of the pre-derived DREME models, the DREME-adapt-pro model, the other DREME-adapt variant, initializes each new fractional reconstruction using the model derived from the immediately preceding fraction (for fraction 1, it uses the model derived from the virtual fraction), which forms a daisy chain and allows the most relevant pre-derived DREME-adapt model to initialize each new reconstruction.

### 2.4 Training strategy of DREME-adapt

The DREME-adapt training strategy is based on DREME (Sec. 2.2)[30], with necessary modifications tailored to the fine-tuning task. In training stage I, the learning rates are reduced to half of those used in the original DREME method for all three substages I-a, I-b and I-c, based on the fine-tuning nature of the adaptive reconstruction approach. The epochs used for I-a and I-b are the same as the original DREME framework, Substage I-a serves as a warm-up for spatial INR ($I_{ref}$) by initializing it from the previous fraction. This initialization naturally passes along the prior information with minimal training/time cost. Substage I-b refines the INR by aligning the digitally reconstructed radiograph from the INR-sampled reference CBCT to the corresponding projection data of the new fraction, to coarsely address the fractional anatomy variations. Another advantage of INR is that it can be partitioned into batches during training when a large coordinate grid is required. We did not apply this strategy in our experiments, since the loss is projection-based and we need to sample the full CBCT volume from the INR to feed into forward projectors to generate DRRs for loss computation. For each training step, we used a projection level-batch, where each projection corresponds to its own DVF and dynamic CBCT. The loss is backpropagated and updates the full INR and resultant $I_{ref}$. In substage I-c, which combines motion model training, DREME-adapt uses only 700 epochs (compared to 1700 epochs for DREME) due to the similarity of motion models between different fractions. In DREME-adapt, the zero-mean score regularization loss $L_{ZMS}$ (Eq. 7) is dropped. Since patient motion can have varying motion patterns and baseline shifts across fractions, to accommodate the fraction-to-fraction baseline shift, the zero-mean score loss was removed during DREME-adapt training.

In training stage II, the learning rates are also reduced to half of those used in the original DREME method. Additionally, the training epochs are significantly reduced. The training substages II-a, II-b and II-c use 200, 100, and 50 epochs, respectively, while all of them require 1000 epochs in the original DREME framework. Similar to DREME, the angular augmentation is added to substages II-b and II-c for DREME-adapt. In contrast, substage II-d is removed from DREME-adapt for acceleration. The DREME model trained on the virtual fraction incorporates angular and motion augmentations, enhancing its generalizability to different motion patterns. Since DREME-adapt is inherently initialized from DREME and further fine-tuned with multi-fractional data, it sufficiently captures motion variations and ensures a robust CNN-based motion encoder; therefore, a separate substage for motion augmentation (II-d) is omitted in DREME-adapt. In summary, DREME-adapt only uses 8.75% of the epochs needed for DREME in the training stage II, leading to much more efficient training. Detailed training setting of the fine-tuning-based DREME-adapt (for both DREME-adapt-vfx and DREME-adapt-pro) can be found in Table 1. The comparison DREME-cs method also used the same training parameter setting as DREME-adapt, without using any ‘warm-start’ initializations. The number of epochs for each substage is determined empirically based on the observation that all loss terms have converged. An example of the evolution of the loss functions during training was provided in the supplementary materials to illustrate the training dynamics of the DREME-adapt.

**Table 1**. The multi-resolution training strategy of DREME-adapt with different loss functions. The symbol ✓ indicates that the corresponding loss function is used in the specific training stage.

| **Training stage** | | **I-a** | **I-b** | **I-c** | **II-a** | **II-b** | **II-c** |
|---|---|---|---|---|---|---|---|
| **Number of epochs** | | 400 | 700 | 700 | 200 | 100 | 50 |
| **Spatial resolution (mm³)** | | | 4×4×4 | | | 2×2×2 | |
| **Spatial INR learning rate** | | 2×10$^{-4}$ | 2×10$^{-5}$ | 5×10$^{-6}$ | 5×10$^{-4}$ | 2×10$^{-4}$ | 5×10$^{-5}$ |
| **Motion encoder learning rate** | | 0. | 0. | 5×10$^{-4}$ | 0. | 0. | 5×10$^{-5}$ |
| **B-spline interpolant learning rate** | | 0. | 0. | 5×10$^{-4}$ | 0. | 0. | 5×10$^{-5}$ |
| **Loss function** | **Weighting factor λ** | | | | | | |
| $L_{sim}^{im}$ | 1. | ✓ | | | ✓ | | |
| $L_{sim}^{proj}$ | 1. | | ✓ | ✓ | | ✓ | ✓ |
| $L_{TV}$ | 2×10$^{-4}$ | | ✓ | ✓ | | ✓ | ✓ |
| $L_{MBC}$ | 1 | | | ✓ | | | ✓ |
| $L_{SC}$ | 1×10$^{3}$ | | | ✓ | | | ✓ |
| $L_{AUG}$ | 1×10$^{-4}$ | | | | | ✓ | ✓ |

2.5 Evaluation datasets and schemes

We evaluated DREME-adapt using the extended cardiac torso (XCAT) digital phantom and a lung patient dataset from the SPARE challenge[7]. The XCAT phantom can simulate different motion scenarios and anatomical variations for quantitative evaluation. The patient study helps to evaluate the clinical application potential of DREME-adapt.

*2.5.1 XCAT simulation study*

We simulated an XCAT phantom to cover the thorax and upper abdomen region, with a volume of 200×200×100 voxels and a 2×2×2 mm$^3$ voxel size. A smooth spherical tumor with an attenuation coefficient of 0.019 mm$^{-1}$ was inserted into the right lung for motion tracking evaluation. The source-to-axis distance (SAD) and source-to-detector-distance (SDD) are set as 1000 mm and 1500 mm, respectively. Seven scenarios (C1-C7) were simulated (Table 2), each with different motion and tumor size variations. In each scenario, six fractional motion curves (one for the virtual scan fraction: VF, and five for CBCT scan fractions: F1-F5) were generated. Each motion curve covers a scan time of 60 seconds. For each VF motion curve, the motion amplitude was randomly selected from a Gaussian distribution with a mean of 20 mm and a standard deviation of 2 mm, and the motion frequency was determined by another Gaussian distribution with a mean of 0.25 Hz and a standard deviation of 0.05 Hz, and both were kept constant from cycle to cycle. The 4D-CT was generated from the first cycle by the XCAT program and then duplicated across all cycles to simulate a dynamic CBCT set with an 11-fps frame rate. For F1-F5 motion curves, inter- and intra-fractional motion frequency variations were introduced. For each curve, the motions frequency gradually increased or decreased, within the range of 0.2 Hz to 0.33 Hz (12 to 20 breath cycles per minute).

For F1-F5, the motion amplitudes were drawn from the same Gaussian distribution as the VF curve, creating inter-fractional amplitude variations. B-spline-based smooth baseline shifts with a random range (lower bound set to -5 mm, and upper bound to 5 mm) were added to these motion curves to mimic the patient breathing baseline shift during the scan. For each scenario, in addition to the fractional motion variations, the tumor was modeled with a shrinking or expanding diameter ranging from 22 mm to 30 mm with a step size of 2 mm, to simulate anatomical changes and to test the robustness of the model and training strategy, though tumor expansion is less likely in clinical practice. Table 2 summarizes the inter- and intra-fractional variations of motion and tumor sizes of the seven XCAT motion scenarios (C1-C7). For each fractional motion curve (F1-F5) of each motion scenario, the 'ground-truth' dynamic CBCT volumes were generated according to the curve at a frame rate of 11 fps, yielding 660 volumes per motion curve. In clinical setups, small rigid misalignment (in addition to anatomical deformations) between 4D-CT and CBCT imaging may occur. To simulate such a potential misalignment and evaluate its effect on DREME-adapt, a rigid translation of vector length $d = 2mm$ is randomly applied to the XCAT volumes along a 3D arbitrary direction between the 4D-CT-based virtual fraction and the other five CBCT fractions. From the virtual and fractional dynamic CBCT sets, we simulated X-ray projections $\boldsymbol{p}$ from them with a rotating gantry (6°/s) in full-fan mode, covering a 360° gantry angle. The X-ray projections were acquired at 11 fps to match with the dynamic CBCT frame rate, so that each dynamic CBCT corresponded to an X-ray projection. Each projection has an image size of 256×192 pixels with a 1.55×1.55 $mm^2$ pixel resolution. The tomographic package ASTRA toolbox was utilized to simulate the cone-beam projections[41], which was also used in the DREME-adapt networks as the cone-beam projector.

**Table 2**. Summary of inter- and intra-fractional variations of motion and tumor sizes of the XCAT motion scenarios. For the motion frequencies, "↓" indicates a decrease from 0.33 Hz to 0.2 Hz, while "↑" represents the opposite. For the tumor size, "↓" indicates a diameter decrease from 30 mm to 22 mm (2 mm per fraction for fractions F1-F5), while "↑" represents the opposite. "-" indicates no change.

| XCAT scenarios | C1 | C2 | C3 | C4 | C5 | C6 | C7 |
| --- | --- | --- | --- | --- | --- | --- | --- |
| Motion frequency variation (inter-fraction and intra-fraction) | - | ↓ | ↓ | ↑ | ↑ | ↓ | ↑ |
| Motion amplitude variation (inter-fraction) | random | random | random | random | random | random | random |
| Baseline shift | random | random | random | random | random | random | random |
| Tumor size (inter-fraction) | - | ↓ | ↑ | ↓ | ↑ | - | - |

The performance of dynamic CBCT reconstruction of different methods was evaluated based on both the image quality of the solved dynamic CBCTs $\boldsymbol{I}(\boldsymbol{x}, p)$ and the accuracy of the lung tumor localization. In detail, the image quality was evaluated via the mean relative error metric (RE):

$$RE = \frac{1}{N_p} \sum_{p} \sqrt{\frac{\sum_{l=1}^{N_{voxel}} \| \boldsymbol{I}(\boldsymbol{x}_l, p) - \boldsymbol{I}^{gt}(\boldsymbol{x}_l, p) \|^2}{\sum_{l=1}^{N_{voxel}} \| \boldsymbol{I}^{gt}(\boldsymbol{x}_l, p) \|^2}}, \tag{11}$$

where $\boldsymbol{I}^{gt}(\boldsymbol{x}, p)$ is the 'ground-truth' CBCT. The similarity index measure (SSIM)[42] and peak signal-to-noise ratio (PSNR)[43] were also evaluated between the reconstructed and 'ground-truth' CBCTs. The tumor localization accuracy was evaluated by contour-based metrics, including the center-of-mass error (COME)[44] ,Dice similarity coefficient (DSC)[45], and 95th percentile Hausdorff Distance (HD95). The 'ground-truth' contours were segmented using a thresholding-based method from the 'ground-truth' XCAT CBCTs. For reconstructed dynamic CBCTs, the tumor was first segmented using the same thresholding-based method from the reference CBCT $\boldsymbol{I}_{ref}$. The segmented tumor was then propagated to other frames using the solved dynamic DVFs and evaluated against the 'ground-truth' contours.

### *2.5.2 Patient study*

The patient dataset includes 7 patients (P1-P7) from the SPARE challenge dataset[7]. Six scans (P1-P4, P6, P7) cover the thoracic region and one (P5) covers the abdominal region. The dataset includes both full-fan (P1-P5) and half-fan (P6, P7) acquisition protocols. For each patient, a simulation 4D-CT set is available. The clinical 4D-CT set only has 10 phases with limited temporal resolution. To generate additional phases and motion states to mimic a dynamic CBCT sequence with higher temporal resolution for the virtual fraction, we employed a motion modeling approach based on volume registration and principal component analysis (PCA)-based DVF decomposition[46]. Specifically, the end-of-exhale phase of 4D-CT was chosen as the reference scan. Non-rigid volume registration, via the Elastix toolbox[47], was then applied between each of the remaining 4D-CT phases and the reference scan to derive the corresponding inter-phase DVFs. The set of DVFs were decomposed as PCA basis vectors with the corresponding coefficients. We chose the first three main PCA basis vectors to represent the motion. We subsequently interpolated the coefficients of these three basis vectors to generate 24 intermediate DVFs. The interpolated DVFs were applied to the reference scan to synthesize 24 more motion states, yielding 34 4D-CT phases in total. We repeated the 4D-CT phases for multiple cycles, creating a total of 340 (full-fan) or 673 (half-fan) volumes to simulate a dynamic CBCT set. Finally, we simulated cone-beam projections from the dynamic CBCT set (one projection from each CBCT), with the gantry angle changing from -90º to 110º (full-fan) or -90º to 270º (half-fan). The projection simulation geometry was matched to the actual projection set of each case (Table 3). In addition, the raw projections generated from 4D-CT-simulated dynamic CBCTs are of different intensity distributions from true cone-beam projections. In clinics, CT and CBCT are acquired by different systems (CT simulator vs. LINAC-equipped flat-panel scanner). Compared with simulation CT, CBCT usually sees more scatter and noise, resulting in different intensity distributions between the two modalities[44]. To minimize their differences, a second-order polynomial regression method was applied to match the intensities of the virtual-fraction cone-beam projections to true projections, providing a simple yet effective solution for model training. Table 3 details the acquisition protocol, number of CBCT fractions (virtual fraction not included), and sizes of projection sets, pixel, and reconstructed CBCTs for each patient case. For each patient case, there is one fraction with both fully-sampled and sparsely-sampled projection sets, with the latter extracted from the former to simulate a faster/lower-dose scan (for the other fractions, only the sparse set is available). For the fraction with both sets available, we used the sparse-sampled projection set to train DREME-adapt, and evaluated the CNN-based motion encoder on the fully-sampled projection set (with the sparse subset removed) to test its real-time motion prediction capability. We arranged the fractional sets in order that the dual-set fraction is always the last fraction. In Table 3, the shown projection number of the test projection sets for real-time prediction is after deducting the sparse subset projection number.

**Table 3.** CBCT imaging and reconstruction parameters for the patient dataset.

| Patient ID | Number of Fractions | Vender | Scan mode | Projection set size for training [a] | Projection set size for real-time prediction [a] | Pixel size ($mm^2$) | SAD[b](mm) / SDD (mm) | Reconstructed CBCT voxels |
|---|---|---|---|---|---|---|---|---|
| P1 | 3 | Elekta | Full fan | 512×512×340 | 512×512×675 | 0.8×0.8 | 1000/1536 | 200×200×100 |
| P2 | 4 | Elekta | Full fan | 512×512×340 | 512×512×665 | 0.8×0.8 | 1000/1536 | 200×200×100 |
| P3 | 3 | Elekta | Full fan | 512×512×340 | 512×512×672 | 0.8×0.8 | 1000/1536 | 200×200×100 |
| P4 | 3 | Elekta | Full fan | 512×512×340 | 512×512×676 | 0.8×0.8 | 1000/1536 | 200×200×100 |
| P5 | 5 | Elekta | Full fan | 512×512×340 | 512×512×667 | 0.8×0.8 | 1000/1536 | 200×200×100 |
| P6 | 2 | Varian | Half fan | 750×1006×(673-679)[c] | 750×1006×1743 | 0.388×0.388 | 1000/1500 | 300×300×102 |
| P7 | 2 | Varian | Half fan | 750×1006×(673-677) | 750×1006×2245 | 0.388×0.388 | 1000/1500 | 300×300×102 |

[a] width (in pixel number) × height (in pixel number) × $N_p$ (number of projections).

[b] SAD: source-to-axis distance. SDD: source-to-detector distance

[c] For the half-fan patients, different fractions have different projection numbers for the sparsely-sampled training set

Since 'ground truth' is unavailable for the patient study, we evaluated the accuracy of the motion estimation by re-projecting the reconstructed CBCTs into DRRs and comparing them with the acquired projections, using motion features tracked by the Amsterdam Shroud (AS) method. Implementation details of the AS method were demonstrated in our previous work[29]. Briefly, for the DRRs and acquired projections, intensity gradients were calculated along the vertical (SI) direction and then added together along the horizonal direction across a region of interest with prominent motion (for instance, the diaphragm). The AS images capture the intensity variations resulting from respiration, and motion traces can be extracted from them to compare the reconstructed motion with that in the original cone-beam projections. To assess the similarity of the extracted motion traces, we computed the localization error[29] and the Pearson correlation coefficient (PCC). As a quantitative evaluation of image quality, we applied the spatial resolution ($\sigma$), defined by the width of diaphragm's edge spread function, as a metric to characterize image sharpness[16]. We first segmented the diaphragm, and then measured three edge profiles orthogonal to the local boundary. Fig. 3(a) shows reconstrued diaphragms by DREME-cs, DREME-adapt-vfx, and DREME-adapt-pro in sagittal view, with their measured edge profiles marked by yellow, red, and blue lines, respectively. The intensity distribution along the edge profile is fitted as a function of edge spread function (ESF) formulated as

$$f(x) = a - \frac{c}{2}\mathrm{erf}\left(\frac{x-r}{\sqrt{2}\sigma_{ESF}}\right), \tag{12}$$

where $a$ is the approximate background attenuation value, $c$ is the contrast between the diaphragm and the background, $x - r$ is the distance to the edge in mm, and $\sigma_{ESF}$ denotes the edge width/spatial resolution in mm. For each fraction, the $\sigma_{ESF}$ is measured for every CBCT frame. Figure 3(b) shows an example of the intensity distribution along the edge profile for all strategies.

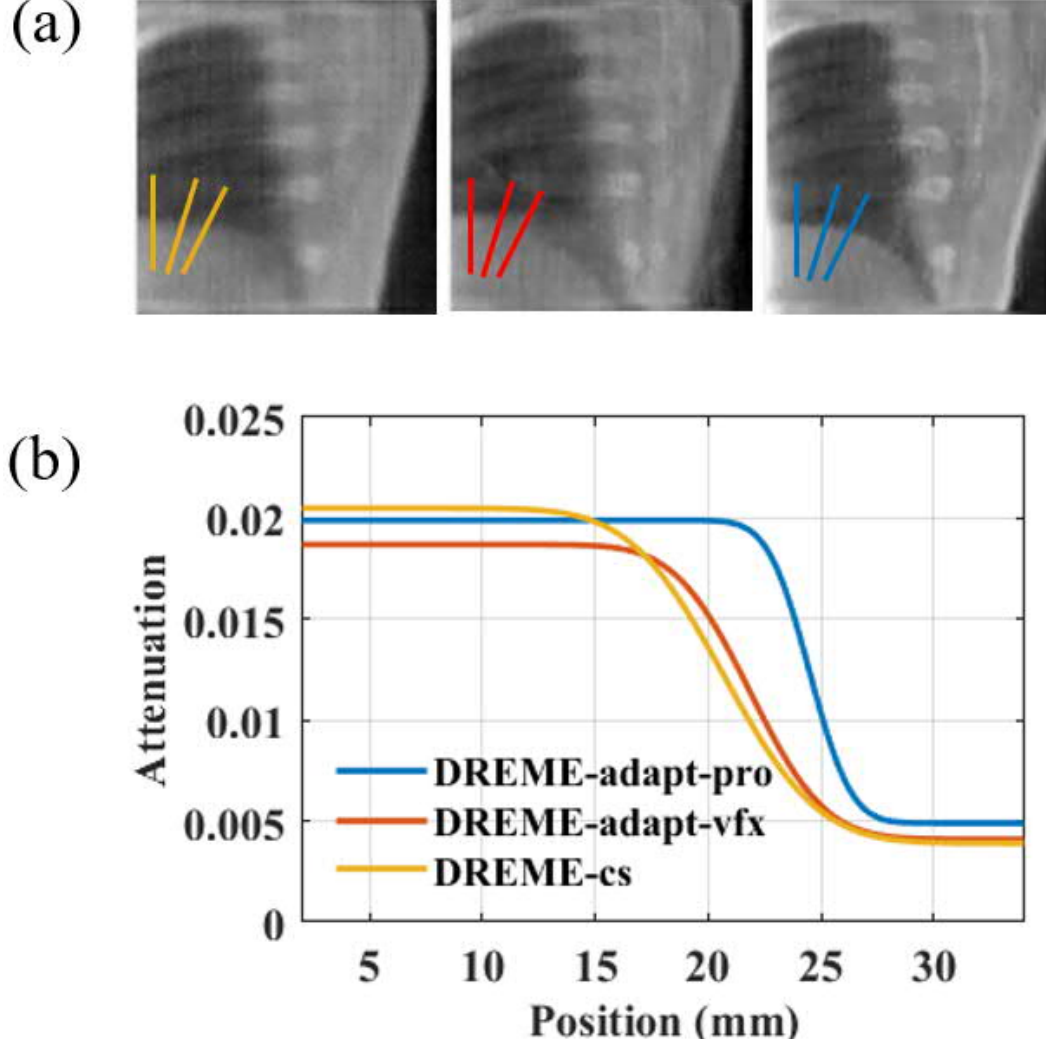

**Figure 3.** (a) Evaluation of the image sharpness. For each frame of reconstructed CBCTs by (left to right) DREME-cs, DREME-adapt-vfx, and DREME-adapt-pro, three intensity profiles (marked as yellow, red and blue lines) of diaphragm are extracted to estimate the edge spread function. (b) Example of the intensity distribution along the edge of the diaphragm using three training strategies.

For both the XCAT simulation study and patient study, we included 3D-CBCT and 4D-CBCT methods for comparison. We used all projections within one fraction to generate a 3D-CBCT reconstruction via the FDK algorithm. For the 4D-CBCT, we first phase-sorted the projections and reconstructed the CBCT for each phase using FDK with additional TV regularization. In the visual comparison, the motion phase of 4D-CBCT most comparable to the displayed frame of DREME-series methods was selected. For the quantitative evaluation, the phase associated with each projection was used to identify its corresponding 4D-CBCT phase for comparison with the 'ground truth'.

## 3. Results

### 3.1 The XCAT study results

Figure 4 compares reconstructed and 'ground-truth' dynamic CBCTs at one frame of fraction 5 (F5) using the different training strategies for all 7 XCAT scenarios (C1-C7), in the coronal (rows a-f) and axial (rows g-l) views. For each view, rows (a, g), (b, h), (c, i), (d, j), (e, k), (f, l) show the reconstruction results from 3D-CBCT, 4D-CBCT, DREME-cs, DREME-adapt-vfx, DREME-adapt-pro, and 'ground-truth', respectively. The tumor and tissue edges were magnified and displayed as insets to highlight their differences across different reconstruction strategies. The reconstruction obtained with 3D-CBCT is affected by motion-induced artifacts and tissue blurring, as no motion model is applied. The resulting images of 4D-CBCT exhibit pronounced artifacts, as this method relies on a limited number of projections for reconstruction. Due to the lack of 'warm-start' knowledge of the anatomy and motion patterns, the results from DREME-cs suffered from residual motion artifacts and incomplete reconstructions as it used the same limited training epoch setting as DREME-adapt (Table 1), resulting in blurry tumors and sternums, and noisy tissues. The DREME-adapt-vfx model was 'warmed-up' with the reconstruction from the virtual fraction. While the tissue appears less noisy, the edges of some structures such as the diaphragm and the tumor are blurred. due to accumulating changes simulated between the virtual fraction and fraction 5 (Table 2). In contrast, DREME-adapt-pro reconstructed the diaphragm and the tumor with sharp edges, leading to minimal motion-induced artifacts, thanks to its progressive, daisy chain design that minimizes the differences between the initialization anatomy/motion model and the to-be-solved ones.

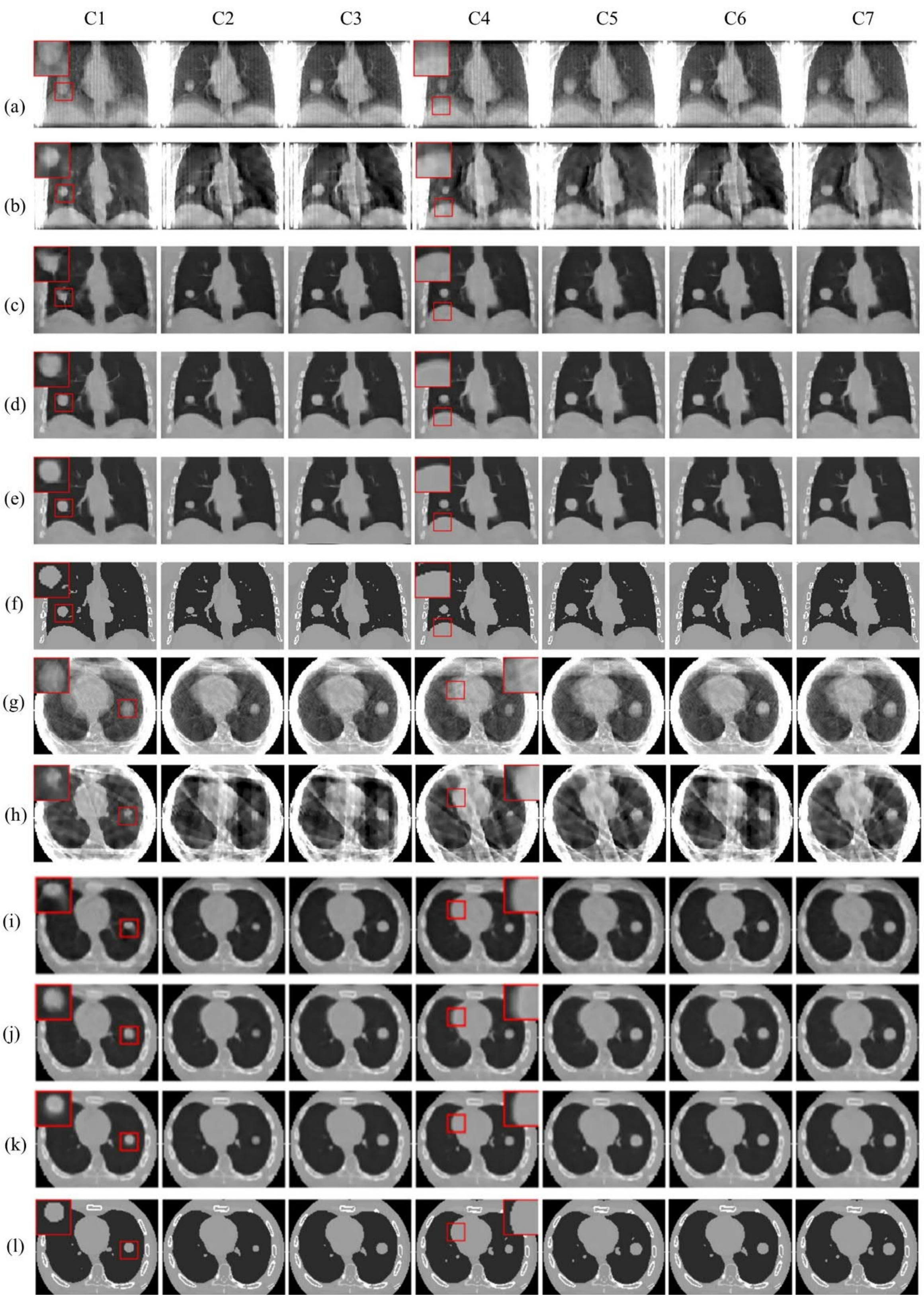

**Figure 4.** Comparison of reconstructed dynamic CBCTs at one frame of fraction 5, for different scenarios (C1-C7) of the XCAT study, between (a, g) 3D-CBCT, (b, h) 4D-CBCT, (c, i) DREME-cs, (d, j) DREME-adapt-vfx, (e, k) DREME-adapt-pro, and (f, l) 'ground truth'. Rows (a-f) and (g-l) show the coronal view and axial view, respectively. DREME-adapt-pro was 'warmed-up' with the anatomy/motion models solved from immediately preceding fractions for fine-tuning, leading to reconstructed dynamic CBCTs with minimal motion-induced artifacts. Red insets highlight the details of tumor and tissue edges as a comparison across different strategies.

To illustrate the fractional results of all three strategies, Figure 5 shows a frame of the dynamic CBCTs solved at different fractions (F1-F5) of scenario C2, in coronal (rows a-d) and axial (rows e-h) views. For each view, rows (a, e), (b, f), (c, g), (d, h) show the reconstruction results from DREME-cs, DREME-adapt-vfx, DREME-adapt-pro, and 'ground-truth', respectively. Note that the same frame may not correspond to the same motion state between F1-F5, due to daily motion variations simulated. In the reconstruction results, the tumor regions were selected and displayed within insets to allow a clear comparison across methods. DREME-cs suffers from blurry diaphragm edges and image noise due to insufficient training. DREME-adapt-vfx suffers from the unclear soft tissue boundaries due to the anatomical variation between the fractions. In contrast, the edge of the diagram was well reconstructed by DREME-adapt-pro from fraction to fraction to capture different motion states, as indicated by the blue arrows. The gradual shrinkage of the tumor size was also successfully captured to match with the simulation (Table 2, C2). The progressive nature of DREME-adapt-pro's training strategy enhances its learning stability, improving its robustness to anatomical and motion pattern variations.

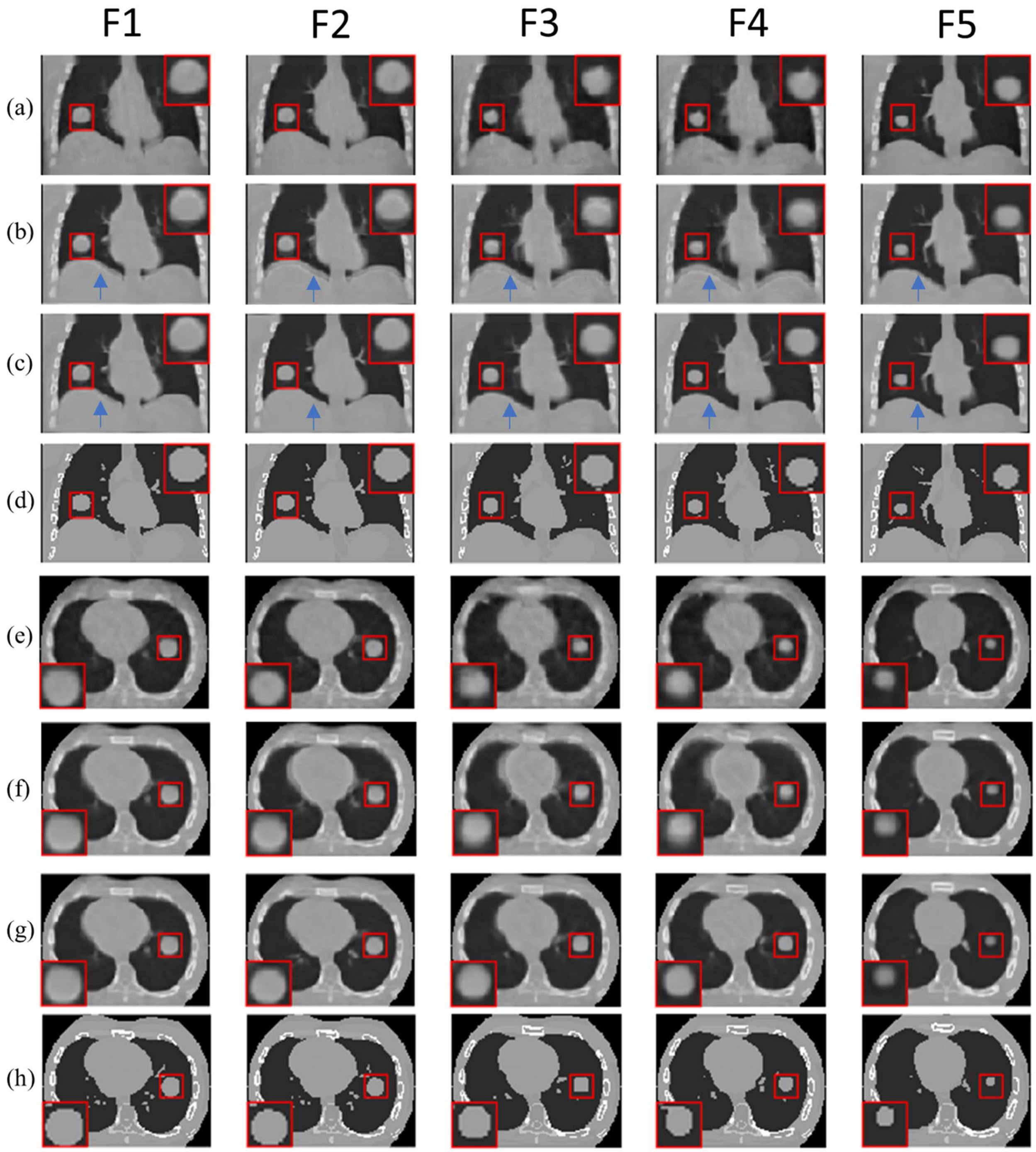

**Figure 5.** Comparison of reconstructed dynamic CBCTs between (a, e) DREME-cs, (b, f) DREME-adapt-vfx, (c, g) DREME-adapt-pro, and (d, h) 'ground truth', across different fractions (F1-F5) for the XCAT scenario C2. Rows (a-d) and (e-h) show the coronal view and axial view, respectively. The tumor of each reconstruction is highlighted within the inset.

The quantitative evaluation of the XCAT simulation study for all training strategies and seven scenarios is summarized in Tables 4 and 5. The results are based on CBCT fraction 5 (F5). 3D-CBCT is not included because of the low performance. In addition, another strategy named DREME-nft (no fine-tuning) is also included for comparison. DREME-nft used the CNN motion encoder of the virtual-fraction DREME model to directly infer the motion fields and the corresponding time-resolved CBCTs from F5 without fine-tuning (Fig. 1), serving as a comparison baseline. Table 4 shows the metrics of dynamic CBCT reconstruction accuracy, including RE, PSNR and SSIM. DREME-adapt-pro outperformed the other strategies in terms of all three metrics. Its chain-based training helps the model to use the most relevant information for 'warming-up', resulting in more accurate anatomy details and reduced motion blur. Table 5 shows the metrics of tumor motion tracking accuracy, including COME, DSC and HD95. Overall, both DREME-adapt-vfx and DREME-adapt-pro achieved sub-voxel COME. And DREME-adapt-pro showed the best results among all strategies. DREME-nft shows better results than DREME-cs in some scenarios, highlighting the benefits of prior knowledge even under anatomy and motion changes. All Wilcoxon signed-rank tests of the six metrics comparing DREME-adapt-pro with each of the other three methods yielded p values $< 10^{-3}$.

**Table 4**. Relative error (RE), structural similarity index measure (SSIM), and peak signal-to-noise ratio (PSNR) of reconstructed dynamic CBCTs of the XCAT simulation study. The results for each XCAT scenario (C1-C7) are presented as the mean and standard deviation (Mean ± SD). All Wilcoxon signed-rank tests of the three metrics comparing DREME-adapt-pro with each of the other three methods yielded p values $< 10^{-3}$.

| XCAT scenarios | **4D-CBCT**<br>RE ↓<br>SSIM ↑<br>PSNR↑ | **DREME-nft**<br>RE ↓<br>SSIM ↑<br>PSNR↑ | **DREME-cs**<br>RE ↓<br>SSIM ↑<br>PSNR↑ | **DREME-adapt-vfx**<br>RE ↓<br>SSIM ↑<br>PSNR↑ | **DREME-adapt-pro**<br>RE ↓<br>SSIM ↑<br>PSNR↑ |
|---|---|---|---|---|---|
| | 1.09±0.03 | 0.23±0.10 | 0.21±0.01 | 0.16±0.01 | **0.15±0.01** |
| C1 | 0.770±0.016 | 0.941±0.033 | 0.933±0.006 | 0.953±0.005 | **0.961±0.006** |
| | 12.63±0.16 | 28.36±3.80 | 28.42±0.69 | 30.42±0.69 | **31.14±0.80** |
| | 1.23±0.02 | 0.16±0.03 | 0.17±0.01 | 0.15±0.01 | **0.14±0.01** |
| C2 | 0.755±0.017 | 0.963±0.011 | 0.954±0.004 | 0.964±0.002 | **0.969±0.004** |
| | 11.56±0.20 | 30.85±1.40 | 30.14±0.56 | 31.51±0.42 | **32.11±0.60** |
| | 1.23±0.02 | 0.16±0.03 | 0.17±0.01 | 0.15±0.01 | **0.14±0.01** |
| C3 | 0.755±0.016 | 0.963±0.011 | 0.953±0.004 | 0.964±0.003 | **0.969±0.004** |
| | 11.55±0.20 | 31.00±1.42 | 30.10±0.56 | 31.44±0.43 | **32.04±0.60** |
| | 1.21±0.01 | 0.18±0.06 | 0.19±0.02 | 0.16±0.02 | **0.14±0.02** |
| C4 | 0.766±0.019 | 0.957±0.019 | 0.943±0.007 | 0.957±0.007 | **0.967±0.008** |
| | 11.73±0.16 | 30.09±2.30 | 29.32±0.79 | 30.75±0.91 | **31.84±1.06** |
| | 1.21±0.02 | 0.17±0.06 | 0.19±0.02 | 0.16±0.02 | **0.14±0.02** |
| C5 | 0.766±0.019 | 0.958±0.019 | 0.943±0.007 | 0.959±0.007 | **0.966±0.008** |
| | 11.75±0.20 | 30.23±2.31 | 29.27±0.83 | 30.96±0.97 | **31.72±1.08** |
| | 1.18±0.01 | 0.16±0.03 | 0.17±0.01 | 0.15±0.01 | **0.14±0.01** |
| C6 | 0.769±0.006 | 0.964±0.011 | 0.953±0.004 | 0.964±0.002 | **0.969±0.003** |
| | 11.96±0.10 | 30.98±1.42 | 30.10±0.56 | 31.42±0.41 | **32.09±0.54** |
| | 1.21±0.012 | 0.17±0.05 | 0.19±0.02 | 0.15±0.02 | **0.14±0.02** |
| C7 | 0.766±0.019 | 0.958±0.019 | 0.943±0.007 | 0.959±0.007 | **0.967±0.008** |
| | 11.73±0.17 | 30.22±2.32 | 29.30±0.83 | 30.99±0.99 | **31.77±1.07** |

**Table 5**. Center-of-mass error (COME), Dice coefficient (DSC) and 95[th] percentile Hausdorff Distance (HD95) as accuracy measures of the solved tumor motion accuracy for the XCAT simulation study. The results for each XCAT scenario (C1-C7) are presented as the mean and standard deviation (Mean ± SD).

All Wilcoxon signed-rank tests of the three metrics comparing DREME-adapt-pro with each of the other three methods yielded p values $< 10^{-3}$.

| XCAT scenarios | **4D-CBCT**<br>COME (mm) ↓<br>DSC ↑<br>HD95 (mm)↓ | **DREME-nft**<br>COME (mm) ↓<br>DSC ↑<br>HD95 (mm)↓ | **DREME-cs**<br>COME (mm) ↓<br>DSC ↑<br>HD95 (mm)↓ | **DREME-adapt-vfx**<br>COME (mm) ↓<br>DSC ↑<br>HD95 (mm)↓ | **DREME-adapt-pro**<br>COME (mm) ↓<br>DSC ↑<br>HD95 (mm)↓ |
|---|---|---|---|---|---|
| C1 | 11.66±5.70<br>0.466±0.229<br>19.23±12.08 | 5.44±7.06<br>0.737±0.287<br>8.29±5.63 | 3.98±2.18<br>0.781±0.059<br>10.85±2.44 | 1.31±0.79<br>0.907±0.039<br>7.57±3.00 | **1.16±0.68**<br>**0.923±0.033**<br>**5.45±0.90** |
| C2 | 7.54±5.77<br>0.565±0.289<br>19.40±12.07 | 1.19±1.63<br>0.589±0.030<br>9.63±1.01 | 1.06±0.58<br>0.904±0.030<br>5.81±0.63 | 0.82±0.48<br>0.899±0.020<br>5.48±0.42 | **0.76±0.33**<br>**0.914±0.025**<br>**5.33±0.54** |
| C3 | 7.52±5.64<br>0.651±0.228<br>8.04±4.89 | 1.10±1.57<br>0.922±0.070<br>5.05±0.97 | 1.26±0.94<br>0.569±0.011<br>5.24±0.57 | 0.87±0.46<br>0.939±0.017<br>4.96±0.37 | **0.79±0.51**<br>**0.942±0.018**<br>**4.69±0.35** |
| C4 | 9.86±7.22<br>0.487±0.304<br>10.11±7.15 | 2.15±3.29<br>0.569±0.078<br>9.63±2.34 | 2.50±1.90<br>0.431±0.025<br>6.62±1.23 | 1.25±1.03<br>0.862±0.052<br>5.09±0.81 | **0.73±0.36**<br>**0.895±0.039**<br>**4.97±0.54** |
| C5 | 9.50±6.77<br>0.575±0.265<br>10.50±5.87 | 2.05±3.19<br>0.879±0.141<br>5.64±2.23 | 2.22±1.73<br>0.571±0.023<br>5.63±0.96 | 1.14±0.94<br>0.924±0.039<br>5.20±0.72 | **1.01±0.79**<br>**0.929±0.032**<br>**4.76±0.59** |
| C6 | 7.20±4.92<br>0.642±0.106<br>17.29±5.74 | 1.13±1.56<br>0.921±0.069<br>5.04±0.96 | 1.24±0.94<br>0.570±0.011<br>5.22±0.59 | 0.87±0.46<br>0.939±0.017<br>5.00±0.37 | **0.72±0.45**<br>**0.947±0.016**<br>**4.83±0.27** |
| C7 | 9.47±6.83<br>0.584±0.263<br>10.94±6.44 | 2.05±3.21<br>0.879±0.142<br>5.64±2.25 | 1.45±1.15<br>0.910±0.050<br>5.68±0.97 | 1.13±0.97<br>0.925±0.040<br>5.21±0.75 | **1.01±0.79**<br>**0.934±0.033**<br>**4.96±0.57** |

Figure 6 compares the solved tumor motion trajectories of different strategies with the 'ground truth' for scenarios C1-C7 of the XCAT study, along both (top) superior-inferior (SI) and (bottom) anterior-posterior (AP) directions. The differences of trajectories are highlighted with insets for each scenario. Overall, DREME-adapt-pro showed more accurate and stable performance. DREME-cs showed a distorted curve, due to the low quality of the motion model. The DREME-adapt-vfx showed relatively smoother curves, which however contain more errors around peaks and valleys than DREME-adapt-pro.

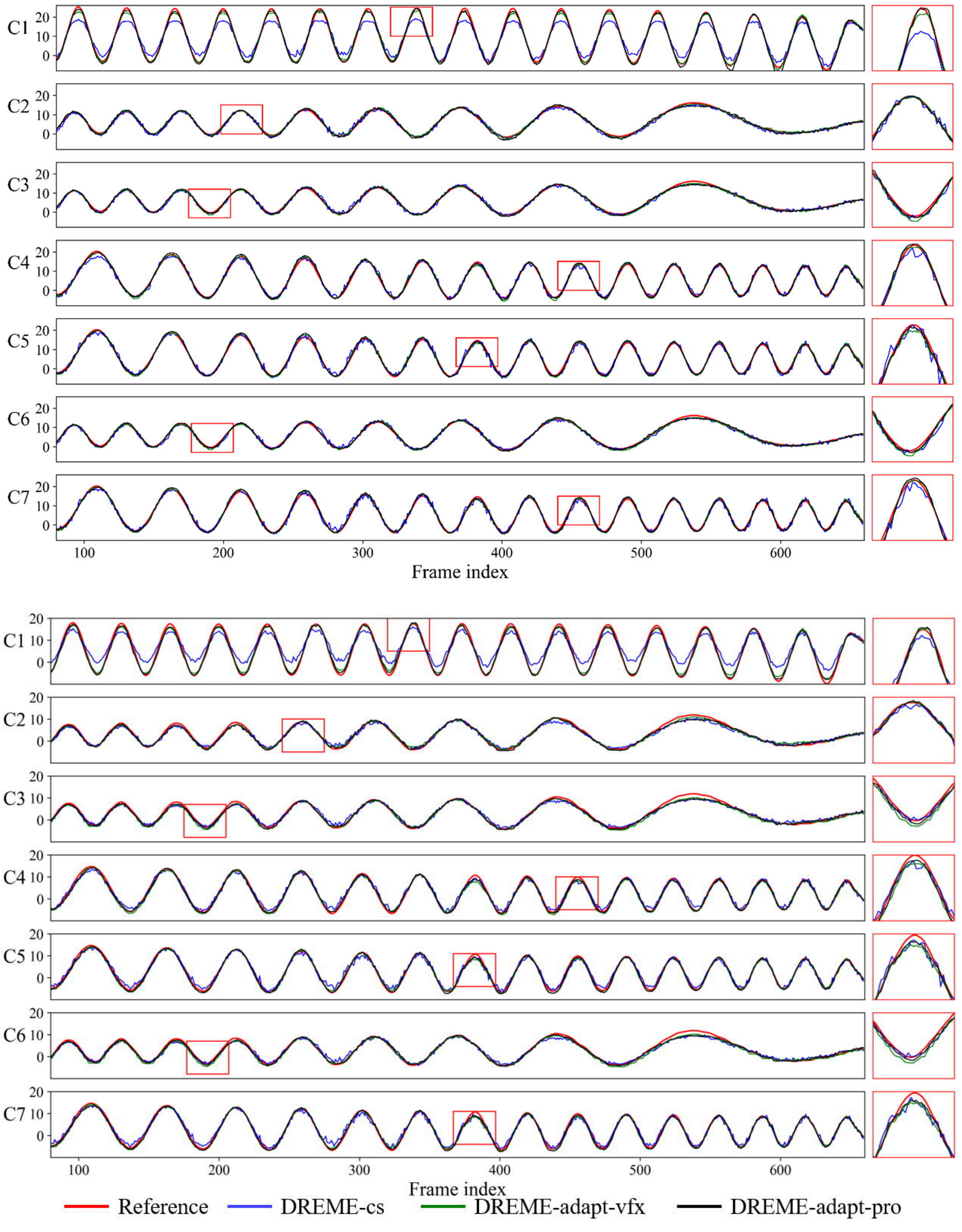

**Figure 6.** Comparison between solved tumor motion trajectories of three training strategies and the 'ground truth' for XCAT study scenarios C1-C7, along the (top) superior-inferior and (bottom) anterior-posterior directions. The differences between trajectories are highlighted with insets.

To demonstrate the effectiveness of the progressive daisy-chain training strategy, the progression of the metrics, (a) RE and (b) COME, during the fractional training of the three strategies is presented in Figure 7 using XCAT scenario C2. In DREME-adapt-pro, both RE and COME consistently exhibit superior performance across all fractions. In scenario C2, the tumor size was set to gradually decrease from F1 to F5, as indicated in Table 2. Consequently, the COME of DREME-cs increases along fractions since the tumor becomes more difficult to track. In contrast, for DREME-adapt-vfx and DREME-adapt-pro, the initialization with prior knowledge allows the models to adapt more effectively to the evolving anatomy.

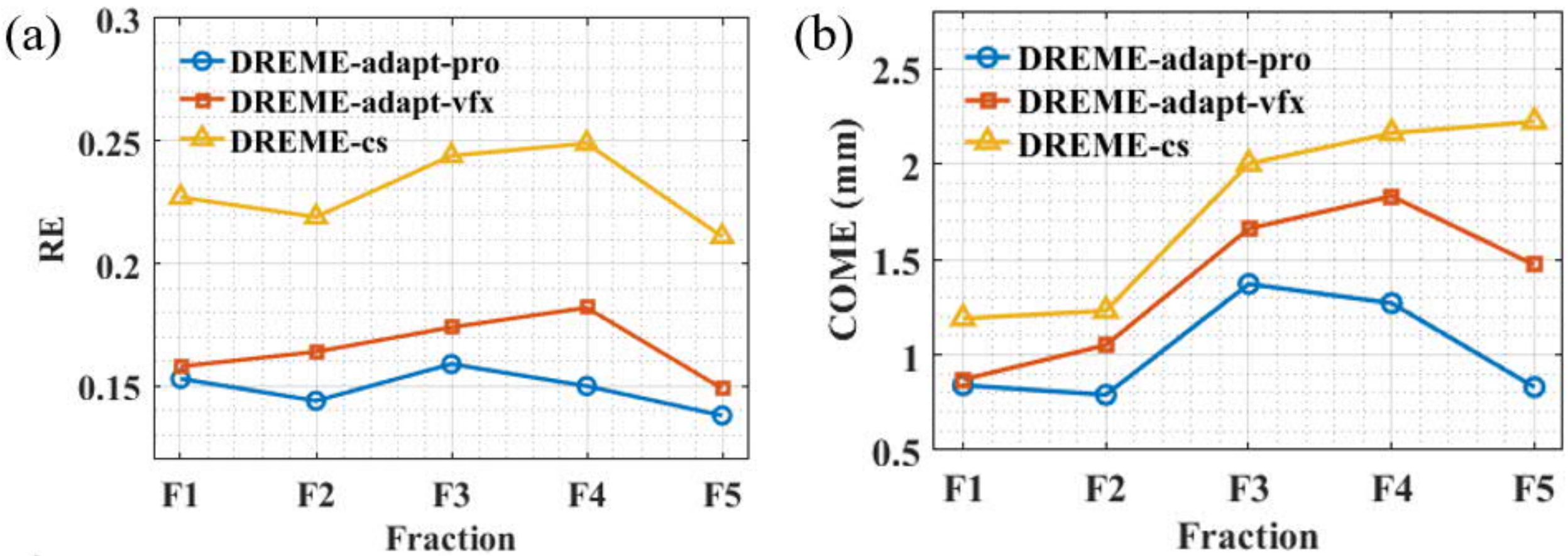

**Figure 7**. (a) RE and (b) COME progression across different fractions (F1-F5) for the XCAT scenario C2 using DREME-cs, DREME-adapt-vfx, and DREME-adapt-pro.

3.2 The patient study results

Figure 8 presents the reconstructed dynamic CBCTs at one time frame of three training strategies and FDK-based methods from the patient study in axial and coronal views, based on the last sparsely-sampled fractional projection set of each case. For clarity, the Fig. 8 is shown in two parts: Part I presents the patients from the full-fan scan, while Part II displays patients from half-fan scans. We also showed the reconstruction results of the original DREME model (with full training) for reference in Fig. 8 (rows: f, l, r). 3D-CBCT (Fig. 8 (a, g, m)) and 4D-CBCT (Fig. 8 (b, h, n)) reconstructions show blurry tissues and are strongly affected by motion- and streak-related artifacts. DREME-adapt-pro (Fig. 8 (e, k, q)) effectively reconstructed varying patient anatomy from projections acquired under different configurations (Table 3), showing its robustness across different imaging protocols. The anatomy reconstructed by DREME-cs (Fig. 8 (c, i, o)) was blurry due to the lack of prior information and insufficient training. Although DREME-adapt-vfx (Fig. 8 (d, j, p)) was trained with a virtual-fraction model for 'warm-start', it still suffers from prominent motion artifacts and blurriness, and tissue hallucinations (for instance, patient P4) due to outdated information within the virtual fraction model. The details of the soft tissue are highlighted in the inset to make clear comparison between different methods.

The localization errors, Pearson correlation coefficients and spatial resolution ($\sigma_{ESF}$) are summarized in Table 6, which are similarly based on the last sparsely-sampled fractional projection set of each case. 3D-CBCT and 4D-CBCT reconstruction results are not included due to low performance. To evaluate the real-time motion inference accuracy of the fine-tuned CNN motion encoders, we used the fully-sampled projection set (Table 3) for direct motion inference and evaluation (the sparse set used during training was excluded from the fully-sampled set, as described in Sec. 2.5.2). The corresponding results are shown in Table 7 to demonstrate the prediction accuracy of different models. DREME-adapt-pro can accurately infer motion from singular X-ray projections, with the predicted motion showing a high correlation (>0.92) and a small mean localization error (2.21 mm) when compared with that directly tracked from the original cone-beam projections using the AS method. Since the localization error was calculated in the projection domain, it would be even smaller after being projected to the 3D space, considering the magnification factor in the projection space (~1.5). DREME-adapt-pro reconstructs images with the clearest anatomical details, achieving a resolution of 1.19 mm as measured at the diaphragm edge. Wilcoxon signed-ranked tests of all metrics between DREME-adapt-pro and other methods were conducted across all patients and all yielded p-values $< 10^{-3}$, confirming the statistical significance of DREME-adapt-pro's higher motion tracking accuracy. Figure 9 compares the predicted and reference SI trajectories of various tracked anatomy for the fully-sampled projection set. The results of DREME-cs were not included due to its poor performance. The inferred motion curves by DREME-adapt-pro and DREME-adapt-vfx matched with the reference, and DREME-adapt-pro had generally better matching, especially at the extrema of the curves as the green insets indicated.

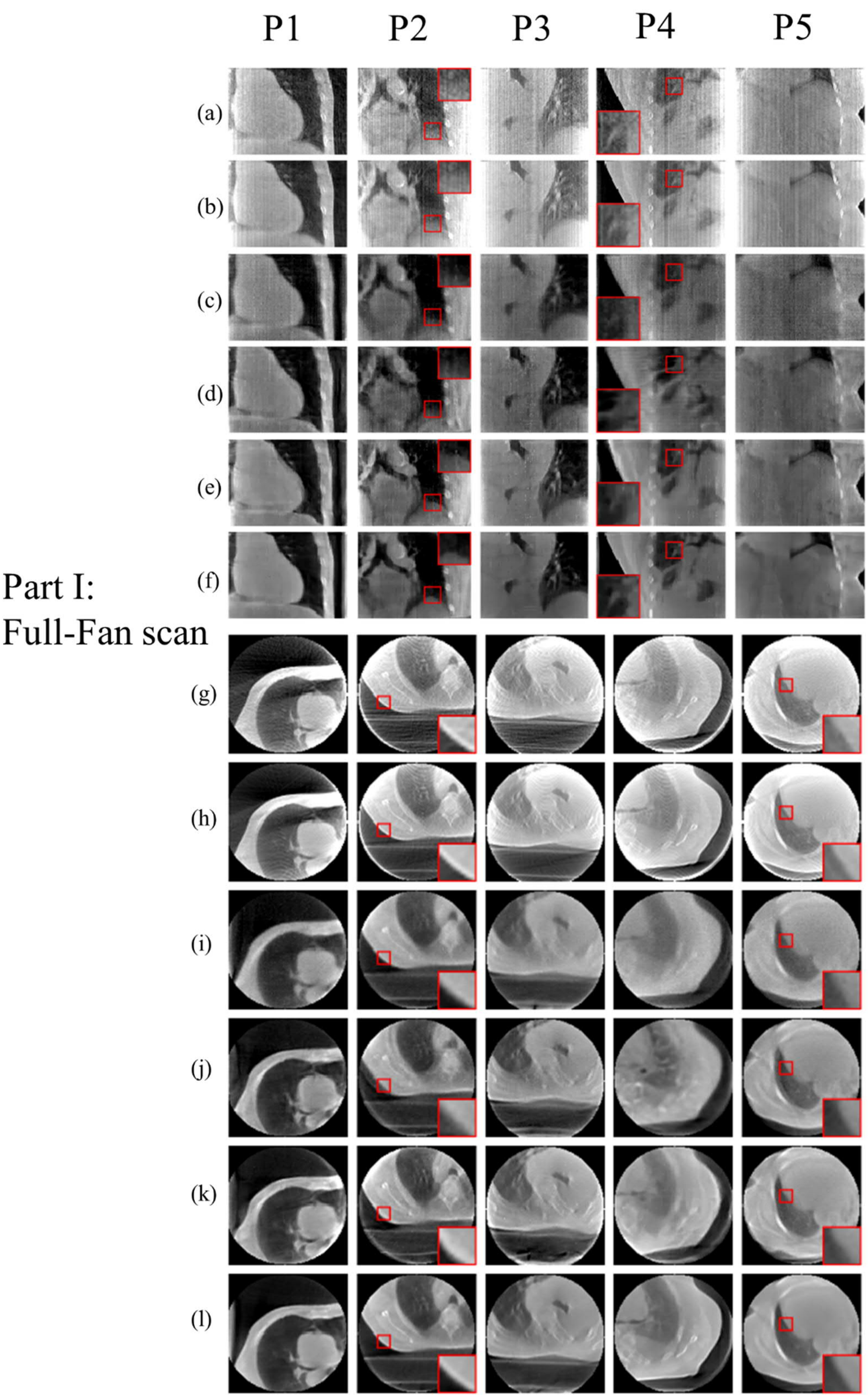

Part II: Half-Fan scan

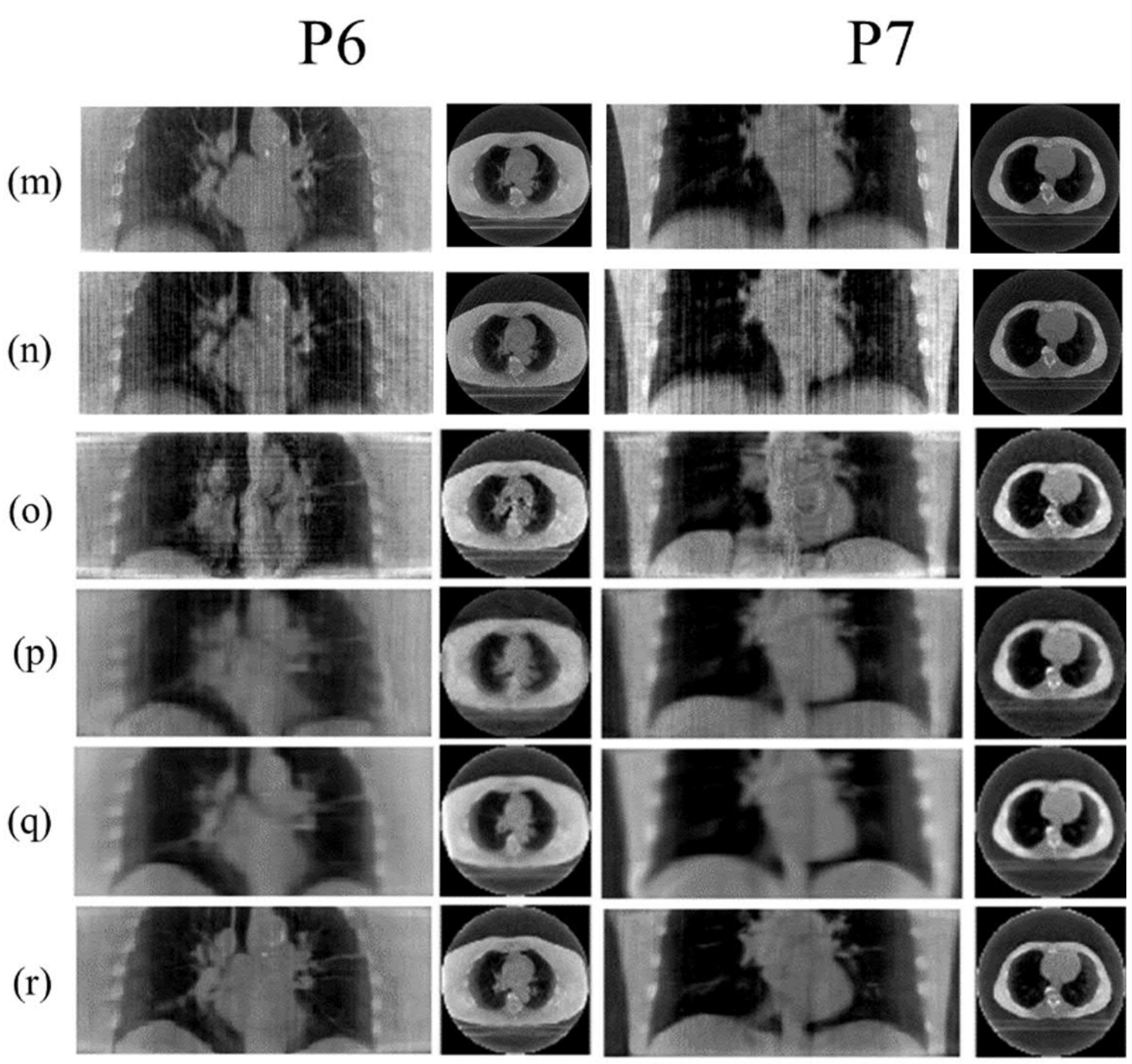

**Figure 8.** Dynamic CBCTs (at one time frame) reconstructed by (a, g, m)3D-CBCT, (b, h, n) 4D-CBCT, (c, i, o) DREME-cs, (d, j, p) DREME-adapt-vfx, (e, k, q) DREME-adapt-pro, and (f, l, r) original full-training DREME for the patient study, shown in coronal and axial views. Figure 8 is divided into two parts: Part I presents the patient from the full-fan scan, while Part II displays patients from half-fan scans. Red insets indicated the tissue edges of each reconstruction. The model training/fine-tuning results were based on the sparsely-sampled projection set of the last fraction for each patient case.

**Table 6.** Pearson correlation coefficient and localization error for the SI motion trajectories extracted from the AS images, and spatial resolution ($\sigma_{ESF}$), of different training strategies for the patient study. The model training/fine-tuning results were based on the sparsely-sampled projection set of the last fraction for each patient case. The localization error results were presented in terms of the mean and standard deviation (Mean ± SD). Wilcoxon signed-ranked tests of all metrics between DREME-adapt-pro and other methods were conducted across all patients and all yielded p-values < $10^{-3}$.

| Patient ID | **DREME-cs**<br>Correlation ↑<br>Localization error(mm) ↓<br>$\sigma_{ESF}$ (mm) ↓ | **DREME-adapt-vfx**<br>Correlation ↑<br>Localization error(mm) ↓<br>$\sigma_{ESF}$ (mm) ↓ | **DREME-adapt-pro**<br>Correlation ↑<br>Localization error(mm) ↓<br>$\sigma_{ESF}$ (mm) ↓ |
|---|---|---|---|
| P1 | 0.942<br>2.72±1.92<br>2.22±0.65 | 0.961<br>2.17±1.68<br>1.38±0.23 | **0.975**<br>**1.77±1.51**<br>**1.36±0.46** |
| P2 | 0.932<br>3.52 ±3.52<br>1.91±0.30 | 0.966<br>2.86±2.18<br>1.46±0.46 | **0.975**<br>**2.59±1.79**<br>**0.85±0.35** |
| P3 | 0.849<br>2.84±2.54 | 0.923<br>2.14±2.08 | **0.932**<br>**1.81±1.89** |
| | 2.09±0.39 | 1.75±0.45 | **1.48±0.42** |
| P4 | 0.936<br>2.81±2.20<br>2.00±0.42 | 0.959<br>2.57±1.72<br>1.95±0.18 | **0.961**<br>**2.47±1.48**<br>**1.71±0.38** |
| P5 | 0.864<br>2.79±3.09<br>2.64±0.37 | 0.942<br>2.37±1.73<br>2.44±0.59 | **0.956**<br>**2.05±1.62**<br>**0.92±0.35** |
| P6 | 0.978<br>2.79±1.80<br>2.00±0.34 | 0.984<br>2.28±1.91<br>1.13±0.22 | **0.989**<br>**1.89±1.56**<br>**1.12±0.18** |
| P7 | 0.872<br>4.74±4.95<br>2.73±2.41 | **0.969**<br>2.82±1.76<br>0.98±0.77 | **0.969**<br>**2.43±2.08**<br>**0.89±0.08** |

**Table 7.** Pearson correlation coefficient and localization error for the SI motion trajectories extracted from the AS images, and spatial resolution ($\sigma_{ESF}$), of different training strategies for the patient study. The results were directly obtained by testing the CNN motion encoder on the fully-sampled projection set of the last fraction for each patient case, with the CNN motion encoder being trained/fine-tuned on another sparsely-sampled projection set of the same fraction, to test its motion prediction accuracy. The results are presented in terms of the mean and standard deviation (Mean ± SD). Wilcoxon signed-ranked tests of all metrics between DREME-adapt-pro and other methods were conducted across all patients and all yielded p-values $< 10^{-3}$.

| Patient ID | **DREME-cs**<br>Correlation ↑<br>Localization error (mm) ↓<br>$\sigma_{ESF}$ (mm) ↓ | **DREME-adapt-vfx**<br>Correlation ↑<br>Localization error (mm) ↓<br>$\sigma_{ESF}$ (mm) ↓ | **DREME-adapt-pro**<br>Correlation ↑<br>Localization error (mm) ↓<br>$\sigma_{ESF}$ (mm) ↓ |
|---|---|---|---|
| P1 | 0.938<br>2.91 ± 2.02<br>2.22±0.67 | 0.954<br>2.34±1.79<br>1.38±0.25 | **0.974**<br>**1.81±1.52**<br>**1.37±0.47** |
| P2 | 0.930<br>3.60 ±3.43<br>1.91±0.30 | 0.965<br>2.90±2.18<br>1.46±0.45 | **0.975**<br>**2.61±1.79**<br>**0.85±0.37** |
| P3 | 0.860<br>2.69±2.47<br>2.09±0.39 | 0.916<br>2.32±2.12<br>1.96±0.57 | **0.926**<br>**2.04±2.00**<br>**1.48±0.42** |
| P4 | 0.930<br>2.87±2.31<br>2.00±0.45 | 0.962<br>2.55±1.72<br>1.95±0.18 | **0.961**<br>**2.49±1.44**<br>**1.71±0.38** |
| P5 | 0.862<br>2.82±3.11<br>2.70±0.35 | 0.933<br>2.45±1.94<br>2.35±0.56 | **0.945**<br>**2.11±1.89**<br>**0.97±0.37** |
| P6 | 0.978<br>2.82±1.78<br>2.00±0.34 | 0.983<br>2.31±1.97<br>1.13±0.22 | **0.988**<br>**1.90±1.53**<br>**1.12±0.19** |
| P7 | 0.866<br>4.82±5.07<br>2.56±2.27 | 0.968<br>2.86±1.77<br>0.97±0.82 | **0.969**<br>**2.53±2.34**<br>**0.89±0.09** |

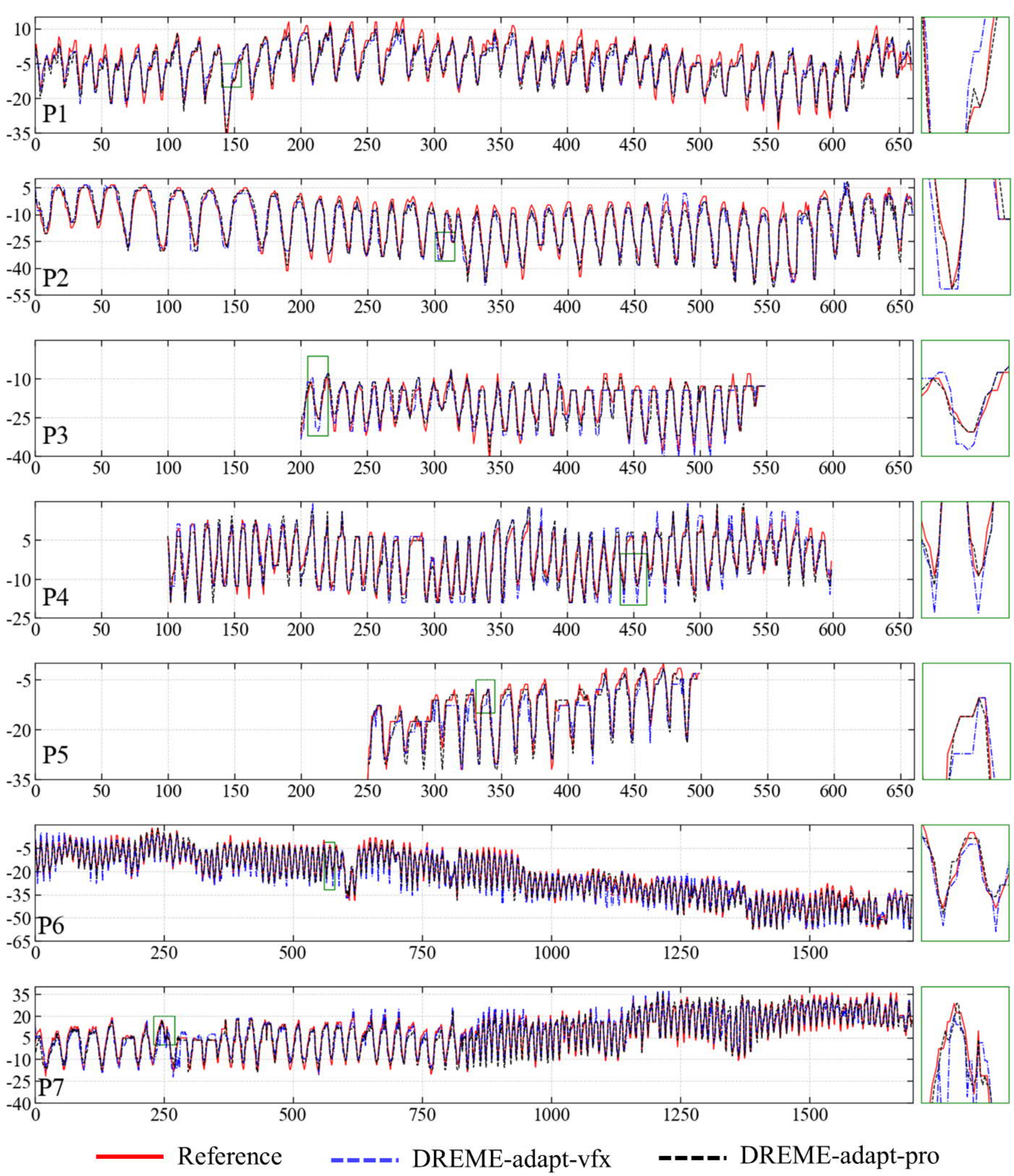

**Figure 9.** Comparison between the reference (red solid line), DREME-adapt-vfx (blue dashed line) and DREME-adapt-pro (black dashed line) motion trajectories extracted using the AS method for the patient study. For some patients, the tracking targets may move out of the field of view, thus only partial trajectories were available from their AS images. The results were directly obtained by testing the CNN motion encoder on the fully-sampled projection set of the last fraction for each patient case, with the CNN motion encoder

being trained/fine-tuned on another sparsely-sampled projection set of the same fraction, to test its motion prediction accuracy.

## 4. Discussion

In this work we proposed an adaptive DREME framework for efficient time-resolved CBCT reconstruction and motion modeling in radiotherapy. The method adopted an adaptive training strategy that incorporates the DREME framework into a prior-guided reconstruction workflow. The patient anatomy and motion models solved from previous fractions are used as prior knowledge to 'warm-start' the reconstruction of following fractions. For fraction 1, due to the lack of prior treatments and CBCT scans, we used patient-specific pre-treatment 4D-CT to create a virtual dynamic CBCT sequence to generate projections to train a virtual DREME model, which serves as the prior. Based on the virtual-fraction DREME model, we evaluated two strategies: DREME-adapt-vfx, which always uses the same virtual-fraction model to 'warm-start' the reconstructions of the following fractions; and DREME-adapt-pro, which uses the virtual-fraction model to initialize the reconstruction of fraction 1, and then uses the resulting 1$^{st}$ fraction model to initialize the reconstruction of fraction 2, forming a progressive, daisy chain-based model. Comparing three training strategies, DREME-cs, DREME-adapt-vfx, and DREME-adapt-pro, DREME-adapt-pro showed the best performance and achieved dynamic image RE of 0.14±0.01, and tumor tracking COME of 0.92±0.62 mm for the XCAT digital phantom study. For the patient study, DREME-adapt-pro achieved a mean localization error of 2.14±1.70 mm, high Pearson correlation coefficients (>0.93) and spatial resolution (1.19±0.32 mm) for the sparsely-sampled training dataset, and a mean localization error of 2.21±1.79 mm, similarly high Pearson correlation coefficients (>0.92) and spatial resolution (1.20±0.33 mm) for the fully-sampled real-time motion inference testing dataset. The inference time for estimating a dynamic CBCT from a projection is approximately 1.5 ms, fulfilling the temporal constraints of real-time imaging[48].

The advantage of DREME-adapt-pro, as compared with the other methods, is the use of the most relevant prior information to initialize the model training via the daisy chain, rendering it more robust to gradual anatomical and motion pattern changes through a treatment course. Note that DREME-adapt does not use the multi-phase motion-resolved 3D volumes of the 4D-CT as "ground-truth" for INR and MBC training because of the following reasons: (i) these 3D volumes often suffer from motion artifacts, which are caused by the combined effects of irregular motion and retrospective sorting (which assumes regular/periodic motion); (ii) anatomical/motion variations between pre-treatment 4D-CT and on-treatment CBCT further limit their reliability as reference datasets for supervised training; and (iii) the 4D-CT is acquired using different hardware from CBCT, and the resulting image quality deviations render direct supervision error-prone. Thus, we only used 4D-CT to train a virtual fraction DREME model, which serves as a start point for adaptation when the projection of a new, real CBCT acquisition becomes available.

Due to the information sharing and communication enabled by the daisy chain scheme, DREME-adapt-pro may benefit from effective 'continual training' and the accumulated training epochs from previous fractions. However, its superior performance cannot be merely attributed to additional training epochs, as anatomical and motion variations between fractions essentially limit the 'accumulation' effect of training epochs. As shown in Fig. 5 and Fig. 7, DREME-adapt-pro outperforms other strategies in the earlier fractions as well, and its reconstruction results are not monotonically improving, suggesting the limited epoch 'accumulation' effect. In comparison, DREME-cs, trained with the same epochs and time, shows motion artifacts and blurriness, as it does not leverage prior knowledge for guidance. Although extending the training time could refine the anatomical details, in realistic clinical scenarios the available training time per fraction is limited, making it necessary to develop strategies like DREME-adapt-pro that perform well under time constraints.

An important aspect of the DREME-adapt-pro framework is the tradeoff between the image quality and the training time. The training time of DREME-adapt-pro is 11 mins, only 15% of the original DREME training. The total training could be divided into two stages: a low-resolution stage, and a high-resolution stage. While the low-resolution stage is efficient in motion model learning, the reference CBCT lacks many detailed features in this stage. The high-resolution stage can refine the details of the reference CBCT image, but the reference CBCT/motion model combined training for a high spatial resolution is more time-consuming. In our full-fan patient case, the 50 epochs of stage II-c took ~2.5 mins, which is 22.7% of the total training time cost. Compared with the original DREME framework, DREME-adapt-pro removes the separate motion augmentation substage to reduce training time, while still maintain a robust CNN motion encoder. This is primarily due to two reasons: (i)DREME-adapt-pro is initialized from a DREME model trained on the virtual fraction, which was trained with both angular and motion augmentations, thereby inheriting its ability to generalize to different motion patterns; (ii) the daisy-chain progressive training strategy allows the model to naturally capture motion variations from the multi-fractional data. Additionally, the iterative DVF self-consistency regulation loss (Eq. 8) could potentially be removed, as it requires an additional 1.7GB of GPU memory and 30 seconds of training time in our experiment. Other lower-cost DVF regularization terms, such as the Jacobian losses[49], can be potentially used instead. However, the accuracy of inverse DVFs could be crucial in clinical applications like dose warping and accumulation[44], in which scenarios the inverse-consistency loss function could be highly desired. The voxel size of CBCT reconstruction is set as $2\times2\times2$ mm³ in this study, as it is sufficient to represent the time-resolved respiratory motion accurately, as such motion is usually smooth. Although the INR can inherently support finer resolutions, the choice of voxel size is constrained by GPU memory requirements, as higher resolution increases both memory consumption and training time. In our case, the DREME-adapt framework requires 24GB of GPU memory for a batch size of 8, under a $2\times2\times2$ mm³ resolution. Recent representation strategy advancements, for instance, Gaussian representation[50], could help to further accelerate the reconstruction speed without compromising on the reconstruction accuracy, while requiring less GPU memory than INR-based methods. The principle of prior-adapted progressive training could be combined with the new Gaussian representation method for faster reconstruction while achieving a finer resolution, which remains to be investigated in the future. The progressive chain-based fine-tuning strategy is also limited by the hash encoder[51]. Because of its discontinuous nature (as compared to Fourier encoding[28]), the fine-tuning of hash encoder is not as smooth as multi-layer perceptron or convolutional layers, leading to salt-and-pepper noise in the CBCT images if insufficient epochs are used for fine-tuning. A solution to this problem is to use another differentiable encoder such as rotary position embedding[52,53], while the corresponding encoding efficiency gain/loss must be evaluated.

Due to the chain structure of DREME-adapt-pro, avoiding model overfitting is important; otherwise, it may reduce the reconstruction accuracy/efficiency of the next fraction. In the current framework, we used a small number (3) of MBCs $\boldsymbol{e}_i(\boldsymbol{x})$ to ensure the low-rank, smooth representation of respiratory motion[36]. We used a $24\times24\times24$ grid of control points for the highest B-spline interpolant to keep a balance between satisfactory motion modeling accuracy and a low risk of overfitting. However, this approach poses challenges in accurately characterizing other motion types, such as the heart beating or random, discontinuous motion caused by coughing or swallowing. To fit the more complex cardiac motion, more MBC levels may be needed[54]. In some cases, the patient may appear significant physiological between scans, we could extend the training epochs of DREME-adapt to capture the anatomical variation. This introduces a trade-off between accuracy and computational time cost. Moreover, the B-spline interpolation could be a computational bottleneck due to the large neighborhood size and substantial memory access requirements. Alternatively, more GPU-optimized techniques such as adaptive sampling-based interpolation[55] or convolution-based method[56] could be investigated in future for efficient interpolation.

Additionally, the training time of the DREME-adapt-pro could be further reduced by more advanced hardware resources, such as newer GPU cards with more powerful processors and improved inter-device memory sharing, as well as efficient fine-tuning techniques. We re-train all parameters of the network in this study. Typical parameter-efficient fine-tuning strategies automatically optimize part of the parameters[57] or modify the network structure by attaching additional layer/parameters while freezing the original parameters[58,59] to use much less GPU memory and achieve faster training, which warrant future investigations to further accelerate DREME-adapt-pro for clinical adoption.

## 5. Conclusion

We proposed a prior-adapted progressive time-resolved CBCT reconstruction framework (DREME-adapt-pro) based on DREME. The proposed method first reconstructs 'virtual' dynamic CBCTs from a pre-treatment 4D-CT and then uses the derived reference anatomy and motion model for sequential fractions in a progressive, adaptive manner. The results demonstrated that the framework yields high-quality motion models with sub-voxel accuracy and a ~85% reduction in training time in both simulation and real patient studies. The method could have broad applications in radiotherapy for daily anatomy tracking and motion-based plan adaptation.

## Acknowledgements

The study was supported by funding from the National Institutes of Health (R01 CA240808, R01 CA258987, R01 CA280135, R01 EB034691), and from Varian Medical Systems. We would like to thank Dr. Paul Segars at Duke University for providing the XCAT phantom for our study.